\documentclass[]{jpp}
\usepackage{amsmath}
\usepackage{graphicx}
\usepackage{xcolor}
\usepackage{natbib}
\usepackage{xpatch}
\usepackage{units}

\usepackage{hyperref}
\hypersetup{
	colorlinks=true,
	breaklinks=true,
	allcolors=blue,
	frenchlinks=true
}

\makeatletter
\xpatchcmd\NAT@citex
{%
	\@citea\NAT@hyper@{%
		\NAT@nmfmt{\NAT@nm}%
		\hyper@natlinkbreak{\NAT@aysep\NAT@spacechar}{\@citeb\@extra@b@citeb}%
		\NAT@date
	}%
}
{%
	\@citea
	\NAT@nmfmt{\NAT@nm}%
	\NAT@aysep\NAT@spacechar
	\NAT@hyper@{\NAT@date}%
}
{}{}
\xpatchcmd\NAT@citex
{%
       \@citea\NAT@hyper@{%
               \NAT@nmfmt{\NAT@nm}%
               \hyper@natlinkbreak{\NAT@spacechar\NAT@@open\if*#1*\else#1\NAT@spacechar\fi}%
               {\@citeb\@extra@b@citeb}%
               \NAT@date
       }%
}
{
       \@citea
       \NAT@nmfmt{\NAT@nm}%
       \NAT@spacechar\NAT@@open\if*#1*\else#1\NAT@spacechar\fi
       \NAT@hyper@{\NAT@date}%
}
{}{}
\makeatother

\usepackage{dsfont}
\usepackage [latin1]{inputenc}

\title{Collisionless relaxation as the origin of anisotropic, non-thermal, and multi-temperature
momentum distributions observed in space plasmas}
\author{Torsten En{\ss}lin\aff{1-4} and Christoph Pfrommer\aff{5}}

\shorttitle{Collisionless relaxation in space plasmas}
\shortauthor{En{\ss}lin and Pfrommer}

\affiliation{
  \aff{1} Max Planck Institute for Astrophysics, Karl-Schwarzschild-Str.\ 1, 85748 Garching, Germany
  \aff{2} Deutsches Zentrum f{\"u}r Astrophysik, Postplatz 1, 02826 G{\"o}rlitz, Germany
  \aff{3} Ludwig-Maximilians-Universit\"at M\"unchen, Geschwister-Scholl-Platz 1, 80539 Munich, Germany
  \aff{4} Excellence Cluster ORIGINS, Boltzmannstr.\ 2, 85748 Garching, Germany
  \aff{5} Leibniz-Institute for Astrophysics Potsdam (AIP), An der Sternwarte 16, 14482, Potsdam, Germany}

\begin{document}
	
\maketitle

\begin{abstract}
Anisotropic, non-thermal, and multi-temperature distributed particle
momenta are commonly observed in collisionless space plasmas, such as the
solar wind. Using Liouville's theorem, we argue that
anisotropic compression or expansion of the plasma,
followed by a relaxation of the resulting anisotropic stress must lead to 
non-equilibrium states that exhibit either anisotropic, non-thermal distribution
functions, different electron and ion temperatures, or a combination of
these effects. We present arguments
showing that a plasma in thermal equilibrium undergoing anisotropic
compression or expansion cannot return to thermal equilibrium in the
absence of particle collisions. Since most astrophysical plasmas are
practically collisionless and experience
significant anisotropic compression or expansion, we expect anisotropic,
non-thermal, and multi-temperature particle distributions to be ubiquitous, in agreement with solar wind measurements.
\end{abstract}


\section{Introduction}

\subsection{Overview}

The momentum space distribution of particles in dilute space plasmas
such as the solar wind is typically not an isotropic Maxwell--Boltzmann
distribution as found in the denser terrestrial gases, but is often better
described by an anisotropic kappa-distribution \citep{1975JGR....80.4181F,1983JGR....88.8893A,marsch2012helios,2019LRSP...16....5V,2020ApJ...892...88B}, while also more complex distributions and such that differ between electrons and ions are observed.
The difference between plasma and gases is most likely due to the different relaxation processes
in the two environments. Molecular collisions in dense gases rapidly establish 
a Maxwell--Boltzmann distribution. By contrast, the observation
of a non-equilibrium momentum distribution in space plasmas clearly signals 
that particle collisions are rare. Collective plasma waves, excited by the free
energy of a non-equilibrium momentum distribution, redistribute the
particle phase space density function and lead to relaxation \citep{2010MNRAS.405..291S}. 

We argue in the following that anisotropies imprinted onto the plasma
should not fully relax to isotropy by such processes. For this argument,
 in a \emph{Gedankenexperiment} we investigate an idealized situation,
which suggests that a full collisionless plasma relaxation does not forget its initial anisotropy. 

\begin{figure}
	\centering
	\includegraphics[width=\linewidth]{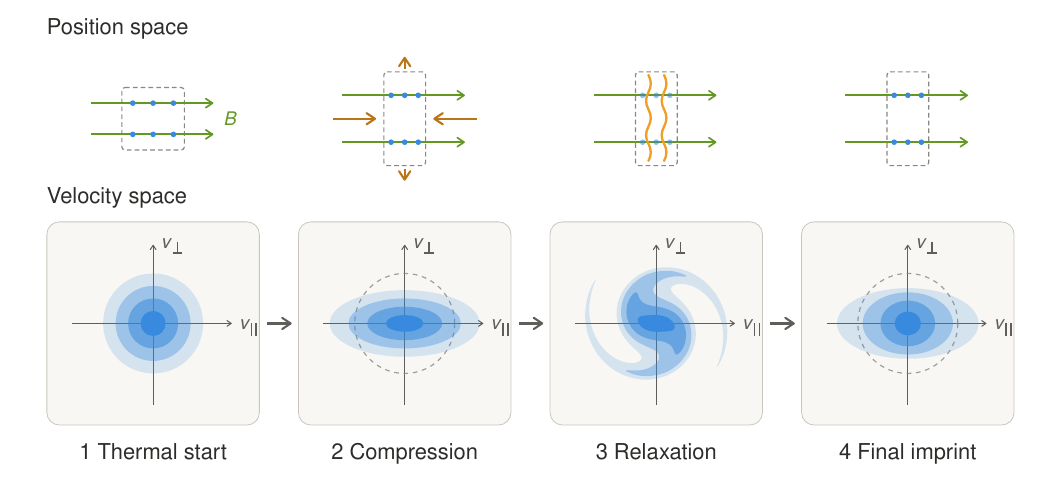}
	\caption{Sketch of the  \emph{Gedankenexperiment} in configuration space (top) and velocity space (bottom). An initially isotropic, eventually thermal plasma gets compressed along magnetic fields and/or expanded perpendicular to those. The resulting anisotropic distribution function becomes unstable to plasma instabilities and this rearranges the phase space density. As particle number, momentum, energy, and all phase space densities need to be conserved in this process, the final distribution function can not be isotropic, as we will show in this work. The figure was generated by the Anthropic language model Faible 5.1 under instruction by the authors.}
	\label{fig:collisionlessrelaxationstagesen}
\end{figure}

Partly it is transferred to higher particle momenta and partly transformed into a temperature difference of electrons and ions.
We argue that real plasmas, such as the solar wind, are often
sufficiently close to this idealized scenario, so that the memory of 
anisotropy created in anisotropic compression or expansion should manifest there as well in a similar fashion. 

The idealized situation we investigate is that of an initially isotropic
plasma volume of practically nearly infinite size without any plasma waves being excited. Let us assume that 
this plasma is compressed or expanded along one of its dimensions, which
gives rise to anisotropy in momentum space. If the imprinted anisotropy creates sufficient
free energy \citep{2017JPlPh..83d7101H}, 
the distribution function will become unstable and plasma
waves are excited \citep{2006ApJ...637..952S,2006AN....327..599S,2010MNRAS.405..291S,yoon_kinetic_2017}.
These reconfigure the momentum space distribution by reducing the
free energy and eventually die out. If during this
process no particles, energy, momentum, or electromagnetic waves are
emitted into the environment of the plasma volume under consideration
or received from there due to the virtual infinite size of our plasma volume, a number of quantities will have to be conserved,
namely energy, momentum, the particle number, and the distribution
of phase space densities \citep{1963PhFl....6..839G,10.1093/mnras/136.1.101,2022JPlPh..88e9201E}. 
In practice that means we are considering regions larger than the damping scale of the waves. 
As a consequence, all energy transferred to the waves need to be returned to the particles in the end and therefore particle energy and phase space densities are conserved.

The latter conservation is due to the fact that the dynamics of a
collisionless plasma is governed by the Vlasov and Maxwells equations,
which form a Hamiltonian system. Hamiltonian systems conserve phase
space densities according to Liouville's theorem.\footnote{Note, 
that the restriction to Hamiltonian systems excludes the investigation of open systems, like 
for example stationary colissionless plasma shock waves. For open systems, phase space volume 
conservation is not enforced, as they exchange entropy with their environment. 
For stationary plasma shock waves this is discussed by 
e.g.~\citet{Paschmann1998AnalysisMF}.} 
We assume that the initial
and final states of our plasma volume are free of macroscopic electromagnetic
fields and that there are no plasma waves before and after the relaxation so 
that the momentum space density function can only be rearranged, but all
phase space densities that existed initially have to be present in
the final state. Using these conserved quantities, we show that the
finally relaxed configuration cannot be an isotropic Gardner distribution \citep{1963PhFl....6..839G,Dodin2005,2017JPlPh..83d7101H,ewart_nastac_schekochihin_2023}, and therefore
also not a Maxwell--Boltzmann distribution. 

Although our proof requires idealized initial and final conditions
to be rigorous, the necessity for imprinted anisotropies to be represented
somewhere in an isolated and collisionless plasma should also hold
for more realistic space plasmas. Temporarily these might be carried
by electromagnetic waves. However, due to the strong coupling of the
charged particles to electric and magnetic fields, a purely electromagnetic
manifestation of the initial anisotropies is therefore necessarily a transient phenomenon.

We argue further that the anisotropies should be imprinted preferentially
onto the outer parts of the momentum space, producing stronger
deviation from Maxwellian distributions at higher particle energies and thereby possibly providing an explanation for the empirically observed kappa-like distributions with strong non-thermal
tails. 

\subsection{Related works}

Non-thermal, anisotropic, and power-law tail particle distributions can occur
 in many physical situations \citep{doi:10.1142/9789812704016-0011,villani2006mathematics,2016PhRvE..93a3110M,Betar_2025}.
In a plasma, sufficiently anisotropic particle distributions are unstable, in particular to the firehose, mirror, and gyrothermal instabilities \citep{chandrasekhar1958stability,southwood1993mirror} as suggested by theoretical and simulation work \citep{2006ApJ...637..952S,2006AN....327..599S,2010MNRAS.405..291S,yoon_kinetic_2017} and observations in the solar wind \citep{maruca2012instability,kasper2013sensitive,klein2017applying,klein2018majority,tong2019statistical,klein2019linear,verscharen2022electron}. At saturation, anisotropy-driven instabilities confine the anisotropy to stay largely within the boundaries of the stability region, which renders the emerging distribution function to be largely isotropic. Because these instabilities draw energy from the anisotropy, they must transfer it elsewhere or convert it into another form. 
For example, there have been speculations by  \citet{2006AN....327..599S} and \citet{2008PhRvL.100h1301S} whether this could lead to explosive
(self-accelerated) magnetic field amplification via firehose and mirror fluctuations.
Because such magnetic fields would be on small scales compared to the scales on which the fluid motions drive anisotropies, and therefore exhibit magnetic curvature, they would try to relax by straightening out, 
and thus drive Alfv\'en and other plasma waves 
on which particles would scatter. Thus, the plasma relaxation process
would remain incomplete in this magnetized state. Here, we are investigating
the state the plasma can finally relax to (if given sufficient time),
and therefore this explosive magnetic field amplification scenario
is excluded from our consideration as a final state.

A central role in our work plays the conservation of the phase space
density or `Casimir' invariants between the compressed and the relaxed
plasma state. 
Such are the volumes of phase space above a certain phase-space density $\varrho$,
\begin{equation}
	V_>(\varrho) := \int d^3x \, d^3p\, \theta(f(\boldsymbol{x},\boldsymbol{p})-\varrho)\, f(\boldsymbol{x},\boldsymbol{p}),
\end{equation}
where $f(\boldsymbol{x},\boldsymbol{p})$ is the six-dimensional (6d) phase space density of a distribution of particles with positions $\boldsymbol{x}$ and momenta $\boldsymbol{p}$ or
-- equivalently -- the (differential) volumes at that density: 
\begin{equation}
	 V(\varrho) := -\frac{d V_>(\varrho)}{d\varrho} = \int d^3x \, d^3p\, \delta(f(\boldsymbol{x},\boldsymbol{p})-\varrho)\, f(\boldsymbol{x},\boldsymbol{p}).
\end{equation}
As the Vlasov equation conserves phase space density (or the corresponding Casimir invariants)  \citet{10.1093/mnras/136.1.101} 
argued that these play a central role for the dynamical equilibrium of such collisionless systems in that 
a specific form of entropy -- later named the Lynden--Bell entropy -- should be maximized.
This is the entropy of the probability density $\mathcal{P}$
that the phase space distribution function $ f(\boldsymbol{x},\boldsymbol{p})$ at some location 
has a certain value $\varrho$, in compact notation 
$\mathcal{P}(f(\boldsymbol{x},\boldsymbol{p})=\varrho )$. 
The Lynden-Bell entropy is therefore
\begin{equation}
	S_\mathrm{LB} = - \int d^3x \, d^3p \, d\varrho  \, \mathcal{P}(f(\boldsymbol{x},\boldsymbol{p})=\varrho ) \ln \mathcal{P}(f(\boldsymbol{x},\boldsymbol{p})=\varrho ).
\end{equation}
 Building on this,
\citet{2022JPlPh..88e9201E,Ewart2025} argued that the final state of an exited, dilute plasma
(excited by a two-stream configuration), since being a collisionless system, should tend to maximize this entropy.

There is, however, an apparent contradiction between these perspectives,
in that the conservation of phase space density (the Casimir invariants)
prohibits any change in Lynden-Bell entropy, as this is also a Casimir invariant.
Thus, the Lynden-Bell entropy cannot be maximized if phase space density is conserved. 

Nevertheless,  \citet{2022JPlPh..88e9201E}
report a change in the phase space distribution function observed in numerical simulations of a two-stream plasma instability. 
In their particular setting, the change of Casimir invariants can be traced back to the coarse graining of the phase space distribution function that their numerical solver performs implicitly. 
This mixes occupied phase space volumes with complete empty ones. Due to the setup of the two-stream instability, in which the intial momentum space distribution is a sum of two delta functions, the occupied and empty regions are direct neighbours in phase space. 
Although the observed phase space mixing in their simulations is a numerical artifact, one could argue that the discreteness of real particle distributions -- the carrier of the idealization of a phase space density -- should lead to similar effects in a real plasma. The assumption of a phase space density fluid made by the Vlasov equation will therefore eventually break down in two-stream situations and distinct phase-space volumes do mix in a similar fashion as reported by \citet{2022JPlPh..88e9201E} and leading to non-thermal, power-law tail distribution functions \citep{ewart_nastac_schekochihin_2023}.

The situation we investigate is, however, very far from that of a two-stream situation, in which occupied and empty phase space volumes are in direct contact. In the adiabatically compressed distribution functions we investigate, neighbouring cells of the phase space are occupied by very similar densities. Thus, any mixing process due to finite resolution (numerical or due to the discrete nature of the particles carrying the density) will hardly change the phase space density. Assuming the Casimir invariants to stay invariant should therefore be a much better assumption in our case, even in the presence of some coarse graining. 

The presence of non-thermal, often kappa-like particle distributions
has been argued to arise from the fact that the plasma entropy
(based on one particle distribution function) should be non-extensive
with respect to the different particle sub-populations. Such non-extensivity arises in the presence of significant correlations between particles, as is naturally expected in systems governed by long-range Coulomb interactions \citep{2018EL....12250001L,2025ApJ...982..169L}.
A manifestation of such correlations is the known effect of Debye
screening of charges in a plasma, meaning that charge carriers of
the same charge are spatially anti-correlated and of the opposite charge
are spatially correlated. Livadiotis and McComas \citep{2018EL....12250001L,2025ApJ...982..169L}
argue that if the addition rule for the single particle entropy
is modified in a specific way, e.g.\ due to two-particle correlations, the
Tsallis entropy appears as the appropriate quantity that describes
an equilibrium, and that this equilibrium is then a kappa-distribution.

Here, we argue that the phase-space configuration to which a collisionless plasma relaxes should be spatially homogeneous, since any inhomogeneities represent free energy and therefore drive further relaxation. Moreover, this configuration should minimize phase-space inhomogeneities, as transport processes, particularly those driven by instabilities, reduce gradients, while conserving particle number, energy, momentum, and Casimir invariants.

The conserved quantities inhibit a complete erosion of phase space
inhomogeneities like anisotropic particle distributions. If the
dynamics smooths out an inhomogeneity in one region of phase space, it
must generate a new one elsewhere, once the field fluctuations
mediating this process, themselves merely a transient form of
inhomogeneity, have subsided.  Thus, the different regions compete for
exporting their inhomogeneity. The momentum space regions which are
fastest in doing so will become most homogeneous, and the ones that
are slower will have to accommodate the inhomogeneity of the others.

In order to get a feeling of how fast phase space inhomogeneities are
erased, the concept of phase space diffusion coefficients is instructive, despite being in contradiction to the phase space density conservation we argue that should hold in our setting.
Quasi-linear theory of particle dynamics in a plasma \citep{1966PhFl....9.2377K,1974JPlPh..12...45L,2002cra..book.....S,2022FrASS...910133B}
predicts that the particle phase space transport coefficients, like
the momentum and pitch angle diffusion coefficients, scale with the
energy density of plasma waves that can resonate with the gyromotion
of particles of a given phase space volume. The level of plasma waves
of a certain kind during relaxation depends on how much free energy
of the corresponding particles is available to excite these waves and how
much anisotropy is present within a given momentum shell. Thus, phase
space regions that harbor a larger density of free energy are more
efficient in exporting it to others. 

For self-excited waves, which emerge during the relaxation process, the diffusion coefficients
become a strongly decreasing function of momentum as soon as the decreasing
part of the particle distribution function is reached. This is simply 
because fewer particles within a given momentum shell carry 
less (free) energy available to excite waves with which they can 
subsequently interact.

Thus, quasi-linear theory predicts that the central areas of the momentum
space will be more efficient in exporting phase space inhomogeneities
than the outer areas. Quasi-linear theory, however, can only be indicative
of what happens, as its prediction of particle diffusion in phase
space is in direct contradiction to the conservation of phase density
due to Liouville's theorem. This only allows for the flow of phase space
density, but not for its mixing.\footnote{Turbulent cascades, however, can transfer power to such small scales that particle collisions, which we ignored here, become unavoidable \citep{nastac2024phase}.} 
More elaborate theories of plasma dynamics,
which do not assume the random phase approximation of quasi-linear
theory, show that the transport of phase space density is more complex and proceeds through the bunching of particle phases \citep{1974JPlPh..12...45L,2025ApJ...979...34L}. 

The outline of this work is as follows: in Section~\ref{sec:Calculations}, we calculate the thermodynamic relaxation in response to anisotropic forcing of  electron--positron and electron--ion plasmas and speculate about the emergence of non-thermal power-law tail towards large momenta. We conclude and discuss further applications of collisionless relaxation to other astrophysical plasmas in Section~\ref{sec:conclusions}.

\section{Anisotropic forcing and relaxation into an equilibrium state\protect\label{sec:Calculations}}

To keep the calculation initially as elementary as possible we first
assume a charge-symmetric plasma, e.g.\ an electron-positron plasma.
This allows us to use particles with a single mass, as the
distribution function should be symmetric with respect to charge exchange.
For an electron--ion plasma, two distribution functions of particles
with different masses need to be considered, which we investigate
afterwards.

\subsection{Notation and conventions}

Nearly every space plasma can be expected to be magnetized to some
level. We assume the magnetic field to be oriented along the $z$-coordinate,
\begin{equation}
\boldsymbol{B}(\boldsymbol{x})=\begin{pmatrix}0\\
0\\
B_{0}
\end{pmatrix},
\end{equation}
and that the adiabatic compression or expansion of the plasma happens
along this direction. 
The latter is an assumption that does not affect
the generality of our argument. 
With adiabatic compression or expansion we refer to any process that is sufficiently slow so that it imprints on all particles in the same way by shrinking or enlarging position space distances between all particles in the magnetic field direction by the same factor, while expanding or compressing them by the same factor in the corresponding direction in momentum space. From an operational point of view, we regard in our Gedankenexperiment processes for which Eqs.~\ref{eq:x-compression}-\ref{eq:end-adiabatic} below hold, and others only to the degree they can be approximated by those formula.

An adiabatic compression or expansion
in another direction will lead to an anisotropy of the distribution
function only w.r.t.\ the magnetic field direction.
The reason is that any action on the particle distribution in one direction perpendicular to the magnetic field (e.g.~the $x$-direction) also acts on the other perpendicular direction to the field (e.g.~the $y$-direction) because of the gyromotion of the particles, which is assumed here to be faster than the adiabatic compression or expansion.\footnote{Situations that violate this assumption certainly exist, but are not part of our Gedankenexperiment.}  
As a consequence, this anisotropy is indistinguishable from that produced by the expansion or compression of an initially isotropic distribution exclusively along the magnetic field.  

As we assume all particles to have the same mass $m$, it is convenient
to express a particle's momentum $\boldsymbol{P}$ in terms of a dimensionless
momentum $\boldsymbol{p}$ and its energy $E$ in units of $m\,c^{2}$, thereby
introducing it's relativistic gamma factor $\gamma$:
\begin{eqnarray}
\boldsymbol{P} & = & \boldsymbol{p}\,m\,c\\
\boldsymbol{p} & = & \begin{pmatrix}p_{x}\\
p_{y}\\
p_{z}
\end{pmatrix}=p\,\begin{pmatrix}\sin\theta\,\cos\varphi\\
\sin\theta\,\sin\varphi\\
\cos\theta
\end{pmatrix}\\
p & := & \sqrt{p_{x}^{2}+p_{y}^{2}+p_{z}^{2}}\\
\mu & := & \cos\theta\\
E(p) & := & \gamma(p)\,m\,c^{2}\\
\gamma(p) & := & \sqrt{1+p^{2}}
\end{eqnarray}
The $z$-compression by a factor $r$ (or expansion by $\nicefrac{1}{r}$
in case $r<1$) transforms the initially isotropic particle distribution in
6d phase space
\begin{equation}
f_{0}(\boldsymbol{x},\boldsymbol{p})
\end{equation}
in the following way: 

\begin{eqnarray}
\label{eq:x-compression}
\boldsymbol{x}=\begin{pmatrix}x\\
y\\
z
\end{pmatrix} & \rightarrow & \boldsymbol{x}'=\begin{pmatrix}x'\\
y'\\
z'
\end{pmatrix}=\begin{pmatrix}x\\
y\\
z/r
\end{pmatrix}=:\boldsymbol{x}'(\boldsymbol{x})\\
\boldsymbol{p}=\begin{pmatrix}p_{x}\\
p_{y}\\
p_{z}
\end{pmatrix} & \rightarrow & \boldsymbol{p}'=\begin{pmatrix}p_{x}'\\
p_{y}'\\
p_{z}'
\end{pmatrix}=\begin{pmatrix}p_{x}\\
p_{y}\\
r\,p_{z}
\end{pmatrix}=:\boldsymbol{p}'(\boldsymbol{p})\\
f_{0}(\boldsymbol{x},\boldsymbol{p}) & \rightarrow & f_{0}'(\boldsymbol{x}',\boldsymbol{p}')\nonumber \\
f_{0}'(\boldsymbol{x}',\boldsymbol{p}') & = & \left.f_{0}(\boldsymbol{x},\boldsymbol{p})\underbrace{\left\Vert \frac{\partial(\boldsymbol{x},\boldsymbol{p})}{\partial(\boldsymbol{x}',\boldsymbol{p}')}\right\Vert }_{=1}\right|_{(\boldsymbol{x},\boldsymbol{p})=(\boldsymbol{x}(\boldsymbol{x}'),\boldsymbol{p}(\boldsymbol{p}'))}\nonumber \\
 & = & f_{0}(\boldsymbol{x}(\boldsymbol{x}'),\boldsymbol{p}(\boldsymbol{p}'))\text{ with}\\
\boldsymbol{x}(\boldsymbol{x}') & = & \begin{pmatrix}x'\\
y'\\
z'\,r
\end{pmatrix}\text{ and }\boldsymbol{p}(\boldsymbol{p}')=\begin{pmatrix}p_{x}'\\
p_{y}'\\
p_{z}'/r
\end{pmatrix}.
\label{eq:end-adiabatic}
\end{eqnarray}

\subsection{Electron--positron plasma}

Now we consider an initially homogeneous and isotropic particle distribution
in 6d phase space:
\begin{equation}
f_{0}(\boldsymbol{x},\boldsymbol{p})=g_{0}(p)
\end{equation}

We assume that $g_{0}(p)$ is a strictly monotonically decreasing function.
The reason is that any phase space density $f_{0}(\boldsymbol{x},\boldsymbol{p})$
that increases with rising $p$ should
lead to plasma instabilities according to Penrose's criterion \citep{1960PhFl....3..258P}, which will erase this rise and yield a flat-top distribution function. 

The adiabatic compression of a volume element in $z$-direction by
a factor $r$ then generates an anisotropic distribution
\begin{eqnarray}
f_{0}'(\boldsymbol{x}',\boldsymbol{p}') & = & g_{0}\left(\sqrt{p'_{x}{}^{2}+p'_{y}{}^{2}+p'_{z}{}^{2}\,r^{-2}}\right).
\end{eqnarray}

Next, we are going to show that collisionless relaxation processes
cannot eliminate an anisotropy in the distribution function. It can
redistribute it along the different momentum levels, but it has to
stay somewhere. We will prove this by contradiction, showing that
if the collisionless plasma relaxation completely re-isotropizes the
pitch angle distribution within our particle pool (or rearranges energies
of electrons and ions separately), and does not create any spatial
structures in the particle density or the electromagnetic fields (after
the relaxation process came to an end), the resulting particle distribution
$f_{1}(\boldsymbol{x},\boldsymbol{p})$ has a different energy compared
to $f_{0}'(\boldsymbol{x},\boldsymbol{p})$, which violates energy
conservation. 

According to the (temporary) assumption of isotropization, the relaxed
distribution has to be of the form of a Gardner distribution  \citep{1963PhFl....6..839G}
\begin{equation}
f_{1}(\boldsymbol{x},\boldsymbol{p})=g_{1}(p),
\end{equation}
with a momentum profile $g_{1}(p)$ that is to be determined from
some constraint: particle number, momentum, energy, and phase-space
density conversation. The latter means that any phase space density
value of the initial distribution has to occupy a similar sized phase-space
volume in the relaxed distribution as the dynamics of collisionless
relaxation fulfills the conditions of Liouville's theorem. We will
also assume that $g_{1}(p)$ is a strictly monotonically decreasing function
and therefore can be inverted. From a plasma physics perspective, this
appears to be a safe assumption, as an increasing occupation number at 
higher momenta is expected to trigger further instabilities 
\citep{1963PhFl....6..839G}.\footnote{Plateaus in the distribution functions, being stable, would prevent the mathematical inversion of the distribution function. Exact plateaus are, however, infinitely unlikely and if they occur can be regarded as the (non-invertible) limit of invertible functions with a tiny negative gradient in the plateau region.} 

First, we set up the condition arising from particle number conversation.
The initial spatial particle density can be expressed as
\begin{eqnarray}
n_{0} & := & \int d^{3}p\:f_{0}(\boldsymbol{x},\boldsymbol{p})
= 
4\pi\,\int_{0}^{\infty}dp\,p^{2}g_{0}(p)
=:
4\pi\,\left\langle p^{2}\right\rangle _{g_{0}},
\end{eqnarray}
where we defined $\left\langle f(p)\right\rangle _{g_{i}}:=\int dp\,g_{i}(p)\,f(p)$
as an $g_i$-weighted integral of $f$, not as its normalized average, \emph{i.e.}
 $\left\langle 1\right\rangle _{g_{i}} = \int dp\,g_{i}(p) =: \overline{g_i} \neq1$
in general.

Similarly, $n_{1}=4\pi\,\left\langle p^{2}\right\rangle _{g_{1}}$.
As the relaxation $f_{o}'\rightarrow f_{1}$ does not change the the
spatial particle density, $n_{1}=n'_{0}=r\,n_{0}$, we find as a necessary
condition 
\begin{eqnarray}
\left\langle p^{2}\right\rangle _{g_{1}} & = & r\,\left\langle p^{2}\right\rangle _{g_{0}},
\end{eqnarray}
a condition that relates $g_1$ to $g_0$,
which we have to verify later.

The plasma momentum is already conserved as
\begin{eqnarray}
\boldsymbol{p}_{1} & = & \int d^{3}p\:f_{1}(\boldsymbol{x},\boldsymbol{p})\,\boldsymbol{p}
=
\int_{0}^{\infty}\!\!\!\!\!\!dp\,p^{3}g_{1}(p)\int_{-1}^{1}\!\!\!\!\!\!d\mu\,\!\!\int_{0}^{2\pi}\!\!\!\!\!\!d\varphi\,\begin{pmatrix}\sin\theta\,\cos\varphi\\
\sin\theta\,\sin\varphi\\
\cos\theta
\end{pmatrix}=\begin{pmatrix}0\\
0\\
0
\end{pmatrix}
\end{eqnarray}
because $g_{1}(p)$ is isotropic in momentum space by construction. Similarly,
\begin{eqnarray}
\boldsymbol{p}_{0}' & = & \int d^{3}p'\:f_{0}'(\boldsymbol{x},\boldsymbol{p}')\,\boldsymbol{p}'=\int d^{3}p\:r\,f_{0}(\boldsymbol{x},\boldsymbol{p})\,\boldsymbol{p}'(\boldsymbol{p})\\
 & = & r\!\!\int_{0}^{\infty}\!\!\!\!\!\!dp\,p^{3}g_{0}(p)\,\int_{-1}^{1}\!\!\!\!\!\!d\mu\,\!\!\int_{0}^{2\pi}\!\!\!\!\!\!d\varphi\,\begin{pmatrix}\sin\theta\,\cos\varphi\\
\sin\theta\,\sin\varphi\\
r\cos\theta
\end{pmatrix}=\begin{pmatrix}0\\
0\\
0
\end{pmatrix}.\nonumber 
\end{eqnarray}
Thus, the plasma momentum is unchanged in the rest frame of the initial plasma, and thus, in any rest frame.

The energy increase (or decrease) happened during the compression,
and the relaxation has to conserve the energy density, thus we have the constraint
\begin{eqnarray}
\varepsilon_{1} & \equiv & \varepsilon'_{0}\text{ with}\\
\varepsilon_{i}^{j} & := & \int d^{3}p\:f_{i}^{j}(\boldsymbol{x},\boldsymbol{p})\,\gamma(p)
\end{eqnarray}
where $\varepsilon_i = \mathcal{E}_i/(m c^2)$ denotes the energy density expressed in units of the rest-mass energy.
The energy density prior to relaxation is
\begin{eqnarray}
\varepsilon'_{0} 
& := & 
\int d^{3}p'\:f'_{0}(\boldsymbol{x}',\boldsymbol{p}')\,\gamma(p')
=
r\int d^{3}p\:f{}_{0}(\boldsymbol{x},\boldsymbol{p})\,\gamma\left(\sqrt{p_{x}{}^{2}+p{}_{y}{}^{2}+p{}_{z}{}^{2}\,r^{2}}\right)
 \label{eq:eps0prime}\\
 & \!\!\!\!=\!\!\!\!\!\! & 2\pi\,r\left\langle p^{2}\!\!\int_{-1}^{1}
 d\mu\,\sqrt{1+\left(1-\mu^{2}\right)p^{2}+\mu^{2}p^{2}r^{2}}\right\rangle _{g_{0}}
 \nonumber \\  
 & \!\!\!\!=\!\!\!\!\!\! &
 2\pi\,r\left\langle p^{2}\!\!\int_{-1}^{1}
 d\mu\,\sqrt{1+p^{2}\left[1-\mu^{2}\left(1-r^{2}\right)\right]}\right\rangle _{g_{0}}\!\!\!\!\!\!.
 \nonumber
\end{eqnarray}
In the non- and ultra-relativistic limits (in the latter case for
compression, $r\ge1)$, this can be further simplified to
\begin{eqnarray}
\varepsilon'_{0} & 
:=
& r\,\begin{cases}
n_{0}+2\pi\,\left\langle p^{4}\right\rangle _{g_{0}}\frac{r^{2}+2}{3} & \text{\!\!non-relativistic,}\\
2\pi\,\left\langle p^{3}\right\rangle _{g_{0}}\!\!\left[r+\frac{\ln\left(r+\sqrt{r^{2}-1}\right)}{\sqrt{r^{2}-1}}\right]\!\! & \!\!\text{ultra-relativistic.}
\end{cases}\nonumber \\
\label{eq:eps0prime-cases}
\end{eqnarray}

The energy density after relaxation is given by
\begin{eqnarray}
\varepsilon_{1} & 
=
& 4\pi\,\int_{0}^{\infty}\!\!\!\!\!\!dp\,p^{2}\sqrt{1+p^{2}}\,g_{1}(p)=4\pi\,\left\langle p^{2}\sqrt{1+p^{2}}\right\rangle _{g_{1}}\!\!\!\!\!\!.
\end{eqnarray}

Once we have constructed the mapping $g_{0}\rightarrow g_{1}$, we
need to check the energy conversation. The mapping $g_{0}\rightarrow g_{1}$
follows from the condition that the phase space density has to be
conserved for every volume element of phase space, while the elements
change locations in momentum space. Thus the amount of phase space
volume $V$ at a given phase space density $\varrho$ has to be conserved
during the relaxation: 
\begin{equation}
V_{1}(\varrho)\equiv V_{0}'(\varrho)\text{ for }\forall\varrho\ge0
\end{equation}
with
\begin{eqnarray}
V_{1}(\varrho) & 
:=
& \int d^{3}p\,\delta\!\left(\varrho-f_{1}(\boldsymbol{x},\boldsymbol{p})\right)=4\pi\int_{0}^{\infty}\!\!\!dp\,p^{2}\,\delta\left(\varrho-g_{1}(p)\right)\nonumber \\
 & \!\!\!\!=\!\!\!\! & 4\pi\int_{0}^{\infty}\!\!\!dp\,p^{2}\,\frac{\delta\left(p-g_{1}^{-1}(\varrho)\right)}{|\partial_{p}g_{1}(p)|} 
 =\frac{4\pi\left[g_{1}^{-1}\left(\varrho\right)\right]^{2}}{\left|\partial_{\cdot}g_{1}\left[g_{1}^{-1}\left(\varrho\right)\right]\right|}
\end{eqnarray}
and 
\begin{eqnarray}
V_{0}'(\varrho) & 
:=
& 2\pi\int_{0}^{\infty}\!\!\!\!\!\!dp\,p^{2}\!\!\!\int_{-1}^{1}\!\!\!\!\!\!d\mu\,\delta\!\left(\varrho-g_{0}\left(p\,\sqrt{1-\mu^{2}\left(1-r^{-2}\right)}\right)\right)\nonumber \\
 & \!\!\!\!=\!\!\!\! & 2\pi\int_{0}^{\infty}\!\!\!\!\!\!dp\,p^{2}\!\!\!\int_{-1}^{1}\!\!\!\!\!\!d\mu\,\frac{\delta\!\left(p-g_{0}^{-1}\left(\varrho\right)/\sqrt{1-\mu^{2}\left(1-r^{-2}\right)}\right)}{\left|\partial_{p}g_{0}\left(p\,\sqrt{1-\mu^{2}\left(1-r^{-2}\right)}\right)\right|}\nonumber \\
 & \!\!\!\!=\!\!\!\! & \frac{2\pi\,\left[g_{0}^{-1}\left(\varrho\right)\right]^{2}}{\left|\partial_{\cdot}g_{0}\left[g_{0}^{-1}\left(\varrho\right)\right]\right|}\,\!\!\!\int_{-1}^{1}\!\!\!\!\!\!d\mu\,\left[1-\mu^{2}\left(1-r^{-2}\right)\right]^{\nicefrac{-3}{2}}
 \nonumber \\
 & \!\!\!\!=\!\!\!\! & \frac{2\pi\,\left[g_{0}^{-1}\left(\varrho\right)\right]^{2}}{\left|\partial_{\cdot}g_{0}\left[g_{0}^{-1}\left(\varrho\right)\right]\right|}\,\!\!\!\int_{-\sqrt{1-r^{-2}}}^{\sqrt{1-r^{-2}}}\!\!\!\!\!\!ds\,\frac{\left[1-s^{2}\right]^{\nicefrac{-3}{2}}}{\sqrt{1-r^{-2}}}\nonumber \\
 & \!\!\!\!=\!\!\!\! & \frac{2\pi\,\left[g_{0}^{-1}\left(\varrho\right)\right]^{2}}{\left|\partial_{\cdot}g_{0}\left[g_{0}^{-1}\left(\varrho\right)\right]\right|}\,\underbrace{\left.\frac{s\,\left[1-s^{2}\right]^{\nicefrac{-1}{2}}}{\sqrt{1-r^{-2}}}\right|_{s=-\sqrt{1-r^{-2}}}^{s=\sqrt{1-r^{-2}}}}_{=2\,r}
 = \frac{4\pi\left[g_{0}^{-1}\left(\varrho\right)\right]^{2}}{\left|\partial_{\cdot}g_{0}\left[g_{0}^{-1}\left(\varrho\right)\right]\right|}r.
\end{eqnarray}
Equating these two volumes $V_1(\varrho)$ and $V_{0}'(\varrho)$ for all $\varrho$ therefore yields, after
a few transformations, the desired relation relation that allows to
calculate $g_{1}$ for a given $g_{0}$ and $r$:
\begin{eqnarray}
\frac{\left[g_{1}^{-1}\left(\varrho\right)\right]^{2}}{\partial_{\cdot}g_{1}\left[g_{1}^{-1}\left(\varrho\right)\right]} & 
=
& \frac{\left[g_{0}^{-1}\left(\varrho\right)\right]^{2}r}{\partial_{\cdot}g_{0}\left[g_{0}^{-1}\left(\varrho\right)\right]}\nonumber \\
\left[g_{1}^{-1}\left(\varrho\right)\right]^{2}\partial_{\varrho}g_{1}^{-1}\left(\varrho\right) & \!\!\!\!=\!\!\!\! & r\left[g_{0}^{-1}\left(\varrho\right)\right]^{2}\partial_{\varrho}g_{0}^{-1}\left(\varrho\right)\nonumber \\
\int_{\varrho}^{\infty}\!\!\!\!\!\!\!\!d\varrho'\left[g_{1}^{-1}\!\!\left(\varrho'\right)\right]^{2}\underbrace{\!\!\partial_{\varrho'}\underbrace{g_{1}^{-1}\!\!\left(\varrho'\right)}_{=:p_{1}(\varrho')}}_{=\frac{dp_{1}}{d\varrho'}} & \!\!\!\!=\!\!\!\! & r\int_{\varrho}^{\infty}\!\!\!\!\!\!\!\!d\varrho'\left[g_{0}^{-1}\!\!\left(\varrho'\right)\right]^{2}\!\!\underbrace{\partial_{\varrho'}\underbrace{g_{0}^{-1}\!\!\left(\varrho'\right)}_{=:p_{0}(\varrho')}}_{=\frac{dp_{0}}{d\varrho'}}\nonumber \\
\int_{p_{1}(\varrho)}^{p_{1}(\infty)}\!\!\!\!dp\,p^{2} & \!\!\!\!=\!\!\!\! & r\int_{p_{0}(\varrho)}^{p_{0}(\infty)}\!\!\!\!dp\,p^{2}\nonumber \\
\frac{1}{3}\left[\underbrace{p_{1}^{3}(\infty)}_{=0}-p_{1}^{3}(\varrho)\right] & \!\!\!\!=\!\!\!\! & \frac{r}{3}\left[\underbrace{p_{0}^{3}(\infty)}_{=0}-p_{0}^{3}(\varrho)\right]\nonumber \\
p_{1}=g_{1}^{-1}(\varrho) & \!\!\!\!=\!\!\!\! & \sqrt[3]{r}\,p_{0}(\varrho)\nonumber \\
\varrho & \!\!\!\!=\!\!\!\! & g_{1}\left(\underbrace{\sqrt[3]{r}\,g_{0}^{-1}(\varrho)}_{:=p}\right)\nonumber \\
g_{1}\left(p\right) & \!\!\!\!=\!\!\!\! & g_{0}\left(p/\sqrt[3]{r}\right).\label{eq:remapping}
\end{eqnarray}
This is the distribution that an isotropic compression would generate after relaxation.
In case there is no compression or expansion, $r=1$, this indeed yields the expected
solution $g_{1}\left(p\right)=g_{0}\left(p\right)$.
Similar constructions of Gardner distributions have been made before \citep[see e.g.,][]{1963PhFl....6..839G,Dodin2005,2017JPlPh..83d7101H,ewart_nastac_schekochihin_2023}. 

Equating these volumes also ensures the conservation of particle number
density during the relaxation as 
\begin{eqnarray}
n_{1} & = & \int d^{3}p\,f_{1}(\boldsymbol{x},\boldsymbol{p})=\int d^{3}p\int_{0}^{\infty}\!\!\!\!\!\!d\varrho\,\varrho\,\delta\!\left(\varrho-f_{1}(\boldsymbol{x},\boldsymbol{p})\right)\nonumber \\
 & = & \int_{0}^{\infty}\!\!\!\!\!\!d\varrho\,\varrho\,V_{1}(\varrho)=\int_{0}^{\infty}\!\!\!\!\!\!d\varrho\,\varrho\,V'_{0}(\varrho)=n_{0}'.
\end{eqnarray}

Energy conservation, however, is not fulfilled as one can
see particular clearly in the non- or ultra-relativistic limits for
$r\ge0$:
\begin{eqnarray}
\varepsilon_{1} & = & 4\pi\,\int_{0}^{\infty}dp\,p^{2}\sqrt{1+p^{2}}\,g_{1}(p)
=
4\pi\,\int_{0}^{\infty}dp\,p^{2}\sqrt{1+p^{2}}\,g_{0}\left(\underbrace{p/\sqrt[3]{r}}_{=p'}\right)\nonumber \\
 & = & 4\pi\,r\,\int_{0}^{\infty}dp'\,p'{}^{2}\sqrt{1+r^{\nicefrac{2}{3}}p'{}^{2}}\,g_{0}\left(p'\right)\nonumber \\
 & \approx & \begin{cases}
r n_{0}+2\pi\,r^{\nicefrac{5}{3}}\left\langle p^{4}\right\rangle _{g_{0}} & \text{non-relativistic}\\
4\pi\,r^{\nicefrac{4}{3}}\left\langle p^{3}\right\rangle _{g_{0}} & \text{ultra-relativistic}
\end{cases}\label{eq:eps1} \\
 & \neq & r\,\begin{cases}
n_{0}+2\pi\,\frac{r^{2}+2}{3}\,\left\langle p^{4}\right\rangle _{g_{0}} & \text{\!\!non-relativistic}\\
2\pi\,\left[r+\frac{\ln\left(r+\sqrt{r^{2}-1}\right)}{\sqrt{r^{2}-1}}\right]\!\!\left\langle p^{3}\right\rangle _{g_{0}}\!\! & \!\!\text{ultra-relativistic}
\end{cases}
\nonumber \\ 
& \approx & 
\varepsilon'_{0}.
\end{eqnarray}
Clearly, at the end of the compression and relaxation cycle all of the injected
energy due to the external compression has to be again in the particle pool, and therefore the energy density before and after relaxation should be identical, and thus $\varepsilon_{1}=\varepsilon'_{0}$
needs to be fulfilled. 

As this is not the case for the isotropic distribution
that conserves volume space density, we can conclude that it is impossible
that an isotropic particle distribution results. Thus, the resulting
particle distribution has to be anisotropic. 

In the language of \citet{2017JPlPh..83d7101H}, we showed that the Gardner state of an isotropic distribution function cannot be reached by collisionless processes only, as it has a different energy than the pre-relaxation state. Some positive \textit{available energy} 
\begin{equation}
	A = \varepsilon_{0}'-\varepsilon_1 \approx 
	\varepsilon_{0} 
	\begin{cases}
	\frac{r^{3}+2\,r}{3}-r^{\nicefrac{5}{3}} &\text{non-relativistic} \\
	\frac{r^2}{2}+\frac{r\,\ln\left(r+\sqrt{r^{2}-1}\right)}{2\,\sqrt{r^{2}-1}} - r^{\nicefrac{4}{3}} &\text{relativistic}
		\end{cases}
	\ge 0
\end{equation}
therefore needs to remain, which gives the distribution function an anisotropic shape. We provide the energy for this adiabatic compression-relaxation cycle for the first time here, to our knowledge.

\subsection{Electron--ion plasma}

We now extend our argument to charge-asymmetric plasmas, such as electron--ion plasmas. We restrict the argument to two species, electrons
with mass $m_{\text{e}}$ and charge $q_{\text{e}}=-e$ and one positive
ion species with mass $m_{\text{i}}$ and charge $q_{i}=Z_{\text{i}}e$,
as the generalization to more species is straight forward. Again,
we assume the initial distribution to be spatial homogeneous, isotropic,
and neutral, implying $n_{\text{e}}=Z_{\text{i}}\,n_{\text{i}}$ and
check whether there is a solution for a spatially homogeneous and isotropic
joint distribution function that factorizes between the species and
that conserves particle number, momentum, and energy.
As before, we assume that plasma processes exist that reconfigure the phase space volumes, without specifying how they do this in detail.   

If there was no interaction between the two species, our single
species argumentation from above would hold. However, electrons and
ions are kinematically different due to their particle masses. 
The two species can interact with each other through collective plasma modes and therefore we cannot ask for energy and phase space volume conservation to
hold individually for each species separately, but only globally for both together. 

The joint initial distribution function $f_{0}$ over the full $6\,N=6\,(N_{\text{e}}+N_{\text{i}})=6\,(1+Z_{\text{i}})\,N_{\text{i}}$-dimensional
phase space, as well as the final $f_{1}$ in the hypothetical case
an isotropization is achieved, is then
\begin{eqnarray}
f_{j}(\boldsymbol{x}_{\text{e},1},\boldsymbol{P}_{\text{e},1},\ldots,\boldsymbol{x}_{\text{i},1},\boldsymbol{P}_{\text{i},1}\ldots,\boldsymbol{x}_{\text{i},N_{\text{i}}},\boldsymbol{P}_{\text{i},N_{\text{i}}}) & =&
\left[\prod_{i_{\text{e}}=1}^{N_{\text{e}}}g_{j}^{\text{e}}(\boldsymbol{P}_{\text{e},i_{\text{e}}})\right]\,\left[\prod_{i_{\text{i}}=1}^{N_{\text{i}}}g_{j}^{\text{i}}(\boldsymbol{P}_{\text{i},i_{\text{i}}})\right]\!\!,\,\,
\label{eq:distribution-function}
\end{eqnarray}
with $j\in\{0,1\}$ labelling the initial and final state and with
restored momenta with units $\boldsymbol{P}_{\text{e}}$ and $\boldsymbol{P}_{\text{i}}$
for electrons and ions, as we have to account for two different masses.
The additional freedom of the two species system is that the relaxation
physics is now allowed to compress the two single particle momentum
space densities differently, exploiting the difference in mass and
charge to treat the two species in different ways. However, due to the product
structure of the phase space density, this allows only 
a single global factor $t=r_{\text{i}}/r$ with which the expansion factor of the ion momentum
space, $r_{\text{i }}$, deviates from the plasma compression factor
$r$, as the expansion of the electron distribution function is then fixed
 to $r_{\text{e}}=r\,t^{\nicefrac{-1}{Z_{\text{i}}}}$. The
exponent expresses that there are $Z_{\text{i}}$ electrons to compensate
for any difference in the momentum expansion from the charge symmetric
value $t=1$, each absorbing a phase space volume compression $r_{\text{e}}$ so that their product-space expands as $r_{\text{e}}^{Z_{\text{i}}}=r^{Z_{\text{i}}}t^{-1}$, which compensates for the $t$-factor in $r_{\text{i }} = r\,t$.\footnote{The point that also the final distributions factorize is essential here.
A violation of this would require that individual particles interact,
which is excluded for a collisionless plasma.}

If we only consider the distribution associated with a single ion species, for which the contribution of a single ion and its $Z_{\text{i}}$ accompanying free electrons to Eq.~\eqref{eq:distribution-function} is
\[
g_{j}^{\text{i}}(\boldsymbol{P}_{\text{i}})
\prod_{i_{\text{e}}=1}^{Z_{\text{i}}}g_{j}^{\text{e}}
(\boldsymbol{P}_{\text{e},i_{\text{e}}})
\]
with $i_{\text{e}}$ being the index running over these electrons, 
before and after the compression-relaxation cycle, we realize that
Eq.\ \eqref{eq:remapping}, $g_{1}\left(p\right)=g_{0}\left(p/\sqrt[3]{r}\right)$,
generalizes to 
\begin{eqnarray}
g_{1}^{\text{i}}(\boldsymbol{P}_{\text{i}})\prod_{i_{\text{e}}=1}^{Z_{\text{i}}}g_{1}^{\text{e}}(\boldsymbol{P}_{\text{e},i_{\text{e}}})
 \label{eq:two-temp-distribution}
 & = & \left[g_{0}^{\text{i}}\left(\frac{\boldsymbol{P}_{\text{i}}}{\sqrt[3]{rt}}\right)t^{-1}\right]\prod_{i_{\text{e}}=1}^{Z_{\text{i}}}
 \left[
 g_{0}^{\text{e}}\left(\frac{\sqrt[3Z_{\text{i}}]{t}}{\sqrt[3]{r}\,}\,\boldsymbol{P}_{\text{e},i_{\text{e}}}\right)t^{\nicefrac{1}{Z_{i}}}\right].
\end{eqnarray}
As electrons and protons are kinematically distinguishable due to their different masses (and charges), we have to allow for the possibility that the two species develop different distribution functions. This gives the collisionless system more freedom to evolve and the possibility to transfer anisotropy into differences between these functions.

Let us first discuss the arguments of the $g$-functions. The ions perceive a compression by a volume factor of $r_\text{i}=rt$, 
and for this reason their three momentum components are each multiplied by $\sqrt[3]{rt}$ in the (assumed) relaxation process, which means that their final distribution function is read off at a momentum that is divided by this factor. The electrons experience a compression of $r_{\text{e}}=r\,t^{-\nicefrac{1}{{Z_{\text{i}}}}}$ and assume corresponding momentum shifts. 
The extra factors $t^{-1}$ and $t^{\nicefrac{1}{Z_{i}}}$ outside the $g$-functions for the ion and electron distribution functions take into account that while
the momentum space distribution is asymmetrically changed, the position
space densities of ions and electrons stay unaffected. Although these
factors compensate each other here, they have to be included in the calculation of the energy densities of the different individual species.
Thus, the corresponding $t$ factors in $g_{1}^{\text{i}}(\boldsymbol{P}_{\text{i}})$ as well as $g_{1}^{\text{e}}(\boldsymbol{P}_{\text{e}})$ have to be read off from the corresponding brackets in Eq.~\eqref{eq:two-temp-distribution}.

It becomes apparent that the additional freedom of having $t\neq1$
allows total energy conservation to be restored, as $t$ affects electron
and ion energies in different ways due to their different masses.
Thus, in principle the dynamics could adapt $t$ in a way that both
electrons and ions could become isotropic, however, for the price
of having now different temperatures. 

\begin{figure}
\includegraphics[width=1\columnwidth]{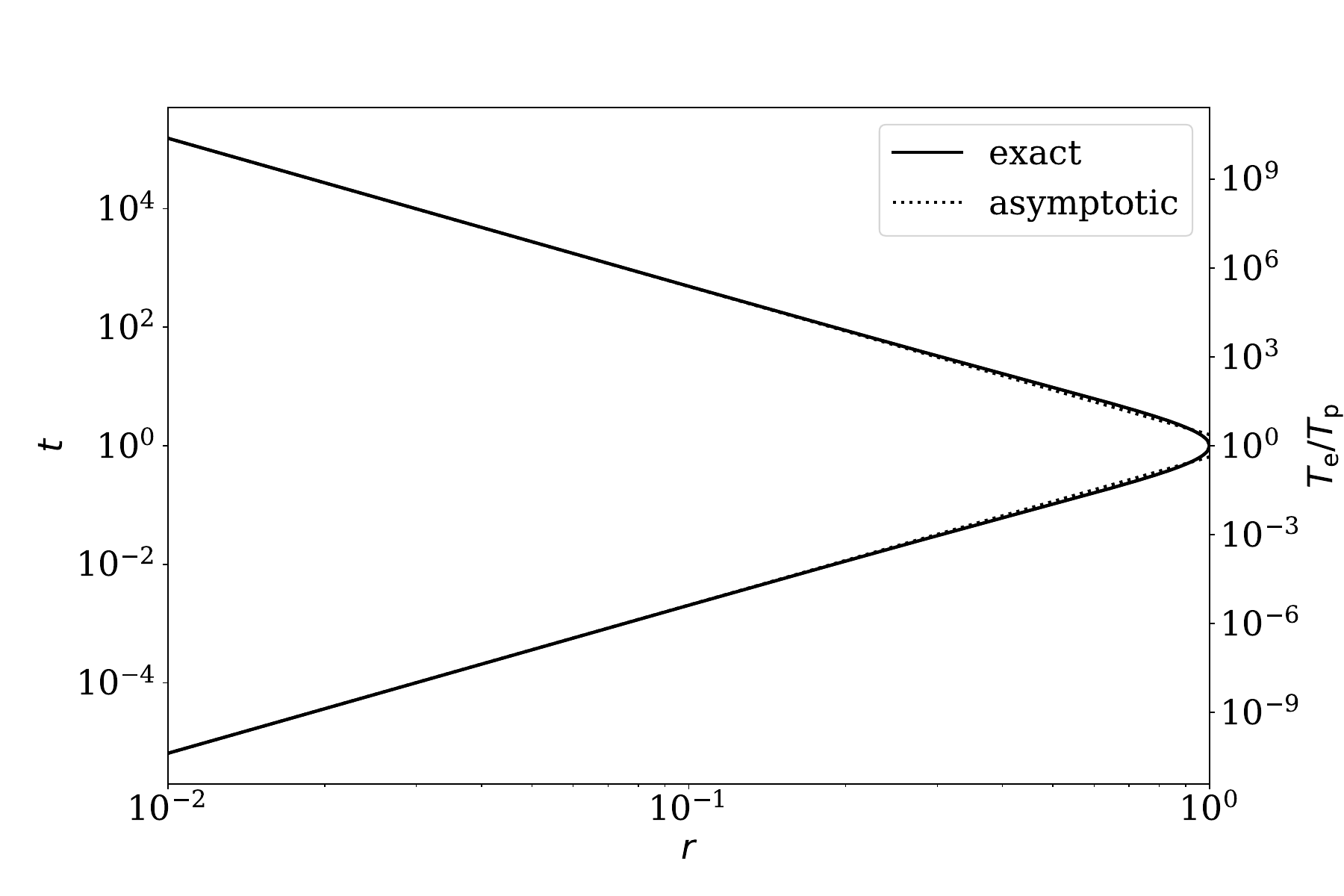}\caption{By phase space and energy density conservation arguments allowed ratio of the momentum space compression ratio (left axis) or temperature
ratio (right axis) for an initially isothermal non-relativistic hydrogen
plasma after expansion by a factor $r^{-1}$ along the magnetic field
direction in case this manages to completely accommodate the initial
anisotropy in an asymmetry of isotropic electron and proton distribution
functions. None of these ratios need to be achieved in reality, as a final state with anisotropic distribution functions is very possible.\protect\label{fig:Ratio-of-the}}
\end{figure}

For illustration, we investigate this for a pure hydrogen plasma ($Z_{\text{i}}=1)$
in the non-relativistic limit. The kinetic energy density of the ions
before and after relaxation is then
\begin{eqnarray}
\varepsilon'{}_{0}^{\text{i}} & = & \frac{\pi}{m_{\text{i}}}\,\frac{r^{2}+2}{3}\left\langle P^{4}\right\rangle _{g_{0}^{\text{i}}}=\frac{r^{2}+2}{3}\varepsilon_{0}^{\text{i}},\\
\varepsilon_{1}^{\text{i}} & = & \frac{\pi\,t^{-1}}{m_{\text{i}}}\,(t\,r)^{\nicefrac{5}{3}}\left\langle P^{4}\right\rangle _{g_{0}^{\text{i}}}=t^{-1}\,(t\,r)^{\nicefrac{5}{3}}\varepsilon_{0}^{\text{i}},
\end{eqnarray}
and that of the electrons is
\begin{eqnarray}
\varepsilon'{}_{0}^{\text{e}} & = & \frac{\pi}{m_{\text{e}}}\,\frac{r^{2}+2}{3}\left\langle P^{4}\right\rangle _{g_{0}^{\text{e}}}=\frac{r^{2}+2}{3}\varepsilon_{0}^{\text{e}},\\
\varepsilon_{1}^{\text{e}} & = & \frac{\pi\,t}{m_{\text{e}}}\,(r/t)^{\nicefrac{5}{3}}\left\langle P^{4}\right\rangle _{g_{0}^{\text{e}}}=t\,(r/t)^{\nicefrac{5}{3}}\varepsilon_{0}^{\text{e}}.
\end{eqnarray}
We refer the reader to Eqs.~\eqref{eq:eps0prime-cases} and \eqref{eq:eps1} for the formulae of the individual energy densities, and Eq.~\eqref{eq:two-temp-distribution} for how $r$ has to be substituted within those formulae to account for the different momentum space expansions of the two species.

Energy conservation during relaxation therefore requires 
\begin{equation}
\frac{r^{2}+2}{3}\left(\varepsilon_{0}^{\text{i}}+\varepsilon_{0}^{\text{e}}\right)=t^{\nicefrac{2}{3}}\,r{}^{\nicefrac{5}{3}}\varepsilon_{0}^{\text{i}}+t^{-\nicefrac{2}{3}}\,r^{\nicefrac{5}{3}}\varepsilon_{0}^{\text{e}}.
\end{equation}
In case of initial equal energy densities in electrons and ions, $\varepsilon_{0}^{\text{e}}=\varepsilon_{0}^{\text{i}}$,
this leads to 
\begin{eqnarray}
\frac{2}{3}\left[r^{\nicefrac{1}{3}}+2\,r^{\nicefrac{-5}{3}}\right] & = & \left[t^{\nicefrac{2}{3}}+t^{-\nicefrac{2}{3}}\right].
\end{eqnarray}
Numerical solutions to this equation as a function of $r$ are shown
in Fig.\ \ref{fig:Ratio-of-the}, as well as the resulting ion-to-electron
energy density (or temperature) ratios. It is interesting to note,
that there exist solutions to this equation only for expansion parallel to
the magnetic field as well as compression perpendicular to it (both being described
by $r\le1)$, but not for parallel compression or perpendicular expansion
($r>1)$ of the plasma. This does not mean that parallel expansion
or perpendicular compression cannot produce different temperatures
for electrons and ions, this simply implies that the anisotropy introduced in this manner cannot be fully captured by an isotropic two-temperature description. Anisotropic two temperature solutions are well possible and probably what nature choses in this regime. 

Asymptotically, we find for the parallel expansion regime
\begin{equation}
t^{\pm1}=\left(\frac{3}{4}\right)^{\nicefrac{3}{2}}r^{\nicefrac{5}{2}}.
\end{equation}
Here the $\pm1$ indicates that if $t$ is a solution, also $\nicefrac{1}{t}$
is a solution, since from pure phase space arguments we cannot determine
whether the electrons or the protons get a larger energy density.
The asymptotic solutions are also shown in Fig.\ \ref{fig:Ratio-of-the}. 

Which solution to the energy and phase space conservation problem
is realized in reality depends on the details of the relaxation process. 
Given that the mass difference between electrons and ions is large, and
therefore also their gyro-radii at the same energy, it is plausible
that they interact with very different wavelengths, and on different
timescales. Thus, to a significant level, their dynamics might be
largely independent. This and the extreme temperature ratios needed
to accommodate the anisotropy via a temperature ratio suggest that
some level of anisotropy remains in the particle distributions. 

A further hint on the answer to this question might be given by plasma compression at shock waves, where more kinetic energy ends up in the ions compared to the electrons 
\citep{treumann2009fundamentals,krasnoselskikh2013dynamic}, thus preferring $t<1$. However, shock waves represent a kinematic regime that is not necessarily covered by our adiabatic compression scenario, and thus give only an indication.

\subsection{Anisotropy in momentum shells}

The previous discussion raises the question of where any residual
anisotropy will reside in momentum space after relaxation.  We give in
the following an heuristic argument why anisotropy is expected to be
mostly present at the larger momenta, leading to a non-thermal tail
for initially thermal distributions.  Although we will argue here a
bit more on the basis of properties of plasma physical processes like
the resonance condition for particle-wave interactions than using
purely conserved quantity arguments, the line of argument stays
qualitatively and speculatively.  The argument given therefore needs
to be confirmed by detailed plasma kinetic simulations.  This is left
for follow up works.

The argument goes as follows: Charged particles interact mostly
resonantly with waves in the relaxation scenario investigated
here,\footnote{We expect non-resonant interactions such as the
\citet{2004MNRAS.353..550B} instability of streaming particles to play
a minor role for comparably moderate particle anisotropies or fluxes
\citep{2021ApJ...908..206S}.} implying that the wavelength and the
gyroradius are of similar scale.  Particles that belong to unstable
regions of phase space -- in our case anisotropically occupied
momentum shells -- excite waves according to the particle-wave
resonance condition that scatter particles of the same and nearby
momentum shells.  This scattering affects largely the pitch angles,
with the effect to establish a more isotropic particle distribution
within the wave driving momentum shell. The removed anisotropy,
however, is not allowed to vanish. It is temporarily present in the
waves, but finally has to be transfered back to the particle
distribution at some different momentum shell.

Thus, it turns out that the different momentum shells of the distribution
try to diminish their anisotropy by exporting it to other shells.
Which of them will win? This is a question of the efficiency with which a
shell exports anisotropy. Initially, all shells receive the same amount
of relative anisotropy during the adiabatic compression or expansion.
The absolute amount of anisotropic energy stored in the various shells
however differs. It is zero in the center of momentum space because
there is no kinetic energy. It rises with increasing momentum until
it reaches a maximum when the particle occupation number starts to
fall off, and then decreases rapidly. Thus, the shell carrying the
largest kinetic energy density will have most anisotropic energy, and
therefore is most efficient in isotropizing itself. It therefore will
transfer the anisotropy to smaller and larger momenta.

At smaller momenta, only a limited amount of this anisotropy will be
absorbed, as energy densities are high (and thus anisotropies are
returned) and no anisotropy can be stored at zero momentum. At higher
momenta, however, there is no such boundary, the resistance against
accepting anisotropy decreases, and every momentum shell can export
its anisotropy to higher shells. 

Thus, the center of momentum space happens to be a leaky box with
respect to anisotropy, exporting this largely to the outer shells,
as far as this is possible within the limits of energy conversation
and plasma wave dynamics.

The transfer of anisotropy from the inner region of momentum space
to the outer region will transfer energy with it, as exciting the
waves by the inner phase space distribution costs these particles
some energy, which is finally absorbed by the outer phase space particles
that scatter off these waves and thereby damp them. 
As the anisotropy
is distributed onto a (typically) fast declining particle density,
larger relative momentum gains for the faster particles are expected,
leading potentially to a power-law-like tail in the distribution function,
as it is given by kappa-like functions.

Thus, we propose that kappa-like distribution functions are often just the result
of anisotropic compression and expansion. For example, the solar wind
plasma expands by a couple orders of magnitude on its journey from
the sun at a radius of about $10^{6}$ km to that of the Earth at around
$10^{8}$ km and further during its journey to the outer solar system. As the radial flow pattern
is shaped by thermal and magnetic pressure, as well as by gravity,
and the magnetic fields are inclined with respect to the flow \citep{1958Parker}, a good
fraction of that expansion will be anisotropic. If the arguments given
above were to hold for the solar wind, a significant non-thermal and
non-isotropic distribution would result, in accordance with measurements. 

In this context the work of \citet{fisk2006common} is relevant. It argues that the origin of the supra-thermal tails in the solar wind is due to acceleration by compressional turbulence. Our phase space based argumentation does not favour any specific set of plasma modes and thus would be consistent with that scenario. Our work, however, adds the possibility that anisotropic plasma expansion (in contrast to compression) can be the origin of the free energy driving turbulence and hence the observed supra-thermal tails.    

\section{Conclusions and outlook}
\label{sec:conclusions}

The conservation of energy, momentum, particle number, and phase space density in collisionless plasmas implies that anisotropic distribution functions, which are naturally arising from anisotropic compression or expansion in astrophysical flows, must be accounted for in some form. The options
of a homogeneously magnetized plasma when being anisotropically compressed
or expanded is to develop
\begin{itemize}
\item a spatially in-homogeneous particle distribution,
\item small scale magnetic field fluctuations,
\item anisotropic momentum distributions,
\item different isotropic distribution functions of the different particle
species, or
\item correlations between individual particles, established in inter-particle
collisions.
\end{itemize}
Nature will probably choose a combination of these different channels.
We argue that for a collisionless plasma, the latter option is hardly
reachable and inhomogeneities in particles and magnetic fields are transient
phenomena, as they represent free energy. Thus, anisotropic and 
species-dependent momentum distributions appear to provide the preferred 
channels for accommodating ultimately the imprinted anisotropy.

Taking the preferred interactions of particles with plasma waves into
account, we expect that the remaining anisotropy is preferentially
transferred to higher energies of the particle distribution function,
increasing deviations from a thermal distribution at these energies.
The idea is that anisotropies on a momentum shell excite plasma waves
of corresponding length scale (given by the particle-wave or wave-wave
resonance conditions). These waves are subject to (collisionless)
damping processes, which transfers energy and anisotropy back to the
particle pool, to the original and to different momentum shells up to
any distance in momentum space that is permitted by particle-wave or
wave-wave scatterings. Thereby, the free energy of anisotropy
propagates through the different shells of momentum space.  As there
is more phase space volume at larger momenta, we expect anisotropies
to accumulate there.

For a charge-symmetric collisionless plasma, like a positron-electron
plasma with both species having the same initial distribution function,
the conservation of anisotropy is exact. For charge-asymmetric collisionless
plasmas, such as electron--ion plasmas, the anisotropy can also be 
accommodated through distinct isotropic distribution functions for each 
species, characterized by different electron and ion temperatures. However,
because electrons and ions preferentially interact with different waves, we
argue that this separation occurs only partially, leaving some of the 
anisotropy in the form of non-thermal distribution functions at higher 
energies.

The qualitative arguments we used to argue about how an anisotropically
compressed or expanded collisionless plasma relaxes need to be verified by detailed
plasma simulation \citep{2023ApJ...948..130T,2026arXiv260109807M}. We hope to be able to present those in future works.

Our considerations on the equilibrium particle distributions emerging from collisionless relaxation are expected to partially carry over from space plasmas to other collisionless astrophysical plasmas in the presence of anisotropic forcing. This might be the case for propagating  shocks as e.g., driven by expanding supernova remnants where the electron-to-ion temperature ratio inferred from a combination of H$\alpha$ line profiles and X-ray spectra is a decreasing function of shock velocity \citep{2007ApJ...654L..69G,2023ApJ...949...50R}, which indicate less efficient electron heating in comparison to kinetic simulations \citep[e.g.,][]{2022ApJ...932...86S}.
Although the assumption of adiabatic compression used here might be violated in shock waves, transient collisionless shocks in closed systems must still obey conservation of phase space density between the pre-shock and post-shock regimes. Thus, our argumentation can be expected to carry over. We leave the question whether they really do to future research.
 
Because of the limitations of current-time numerical simulation regarding the dimensionality, run time and resolution of the kinetic simulations, our considerations may enable identifying the missing pieces that would have to be incorporated first in future simulations. 

Moreover, weakly collisionless plasmas such as in galaxy clusters are prone to anisotropic forcing following cluster mergers and the excitation of intracluster turbulence as demonstrated in recent cosmological simulations \citep{2026arXiv260106250P} and inferred from the observed X-ray line broadening \citep{2025ApJ...982L...5X}. If the assumed particle distribution is not a superposition of thermal Maxwellians but instead composed of kappa distributions, this could bias the inference of the turbulent velocity distribution. However, care must be taken when interpreting spectra containing non-thermal electrons, as it must be demonstrated that the observed spectrum is not simply the superposition of two distinct plasmas with different temperatures and emission measures \citep{2015ApJ...809..178H,2019LRSP...16....5V}.

\section*{Funding}
CP and TE acknowledge support by the European Research Council under ERC-AdG grant PICOGAL-101019746 and under ERC-SyG grant ERC SyG mw-atlas 101166905, respectively. TE furthermore gratefully acknowledge the financial support provided by the German Federal Ministry of Research, Technology and Space (BMFTR) in the framework of the Knowledge creates perspectives for the region!, for the project StStG -- DZA -- Aufbauphase: Deutsches Zentrum f\"ur Astrophysik, Gro{\ss}forschungszentrum in der s\"achsischen Lausitz: Aufbauphase 2026, grant number 03WSP1746.

\section*{Declaration of Interests}
The authors report no conflict of interest.

\section*{Data Availability}
The data underlying this article will be shared on reasonable request to the corresponding author.

\section*{Author ORCID}

T.\ En{\ss}lin, \href{https://orcid.org/0000-0001-5246-1624}{https://orcid.org/0000-0001-5246-1624}

C.\ Pfrommer, \href{https://orcid.org/0000-0002-7275-3998}{https://orcid.org/0000-0002-7275-3998}



\bibliographystyle{jpp}
\bibliography{bibliography}

\begin{thebibliography}{54}
\expandafter\ifx\csname natexlab\endcsname\relax\def\natexlab#1{#1}\fi
\def\au#1{#1} \def\ed#1{#1} \def\yr#1{#1}\def\at#1{#1}\def\jt#1{\textit{#1}}
  \def\bt#1{#1}\def\bvol#1{\textbf{#1}} \def\vol#1{#1} \def\pg#1{#1}
  \def\publ#1{#1}\def\arxiv#1{#1}\def\org#1{#1}\def\st#1{\textit{#1}}

\bibitem[{Armstrong} {\em et~al.\/}(1983){Armstrong}, {Paonessa}, {Bell} \&
  {Krimigis}]{1983JGR....88.8893A}
{\sc \au{{Armstrong}, T.~P.}, \au{{Paonessa}, M.~T.}, \au{{Bell}, II, E.~V.} \&
  \au{{Krimigis}, S.~M.}} \yr{1983}  \at{{Voyager Observations of Saturnian Ion
  and Electron Phase Space Densities}}.  \jt{\jgr}  \bvol{88}~(A11),
  \pg{8893--8904}.

\bibitem[{Bell}(2004)]{2004MNRAS.353..550B}
{\sc \au{{Bell}, A.~R.}} \yr{2004}  \at{{Turbulent amplification of magnetic
  field and diffusive shock acceleration of cosmic rays}}.  \jt{\mnras}
  \bvol{353}~(2),  \pg{550--558}.

\bibitem[{Ber{\v{c}}i{\v{c}}} {\em et~al.\/}(2020){Ber{\v{c}}i{\v{c}}},
  {Larson}, {Whittlesey}, {Maksimovi{\'c}}, {Badman}, {Landi}, {Matteini},
  {Bale}, {Bonnell}, {Case}, {Dudok de Wit}, {Goetz}, {Harvey}, {Kasper},
  {Korreck}, {Livi}, {MacDowall}, {Malaspina}, {Pulupa} \&
  {Stevens}]{2020ApJ...892...88B}
{\sc \au{{Ber{\v{c}}i{\v{c}}}, Laura}, \au{{Larson}, Davin}, \au{{Whittlesey},
  Phyllis}, \au{{Maksimovi{\'c}}, Milan}, \au{{Badman}, Samuel~T.},
  \au{{Landi}, Simone}, \au{{Matteini}, Lorenzo}, \au{{Bale}, Stuart.~D.},
  \au{{Bonnell}, John~W.}, \au{{Case}, Anthony~W.}, \au{{Dudok de Wit},
  Thierry}, \au{{Goetz}, Keith}, \au{{Harvey}, Peter~R.}, \au{{Kasper},
  Justin~C.}, \au{{Korreck}, Kelly~E.}, \au{{Livi}, Roberto}, \au{{MacDowall},
  Robert~J.}, \au{{Malaspina}, David~M.}, \au{{Pulupa}, Marc} \& \au{{Stevens},
  Michael~L.}} \yr{2020}  \at{{Coronal Electron Temperature Inferred from the
  Strahl Electrons in the Inner Heliosphere: Parker Solar Probe and Helios
  Observations}}.  \jt{\apj}  \bvol{892}~(2),  \pg{88},  \arxiv{arXiv:
  2003.04016}.

\bibitem[Betar \& Del~Sarto(2025)]{Betar_2025}
{\sc \au{Betar, Homam} \& \au{Del~Sarto, Daniele}} \yr{2025}  \at{Microscopic
  current sheets and fast tearing modes in plasma turbulence}.  \jt{The
  Astrophysical Journal}  \bvol{990}~(1),  \pg{28}.

\bibitem[{Brizard} \& {Chan}(2022)]{2022FrASS...910133B}
{\sc \au{{Brizard}, Alain~J.} \& \au{{Chan}, Anthony~A.}} \yr{2022}
  \at{{Hamiltonian formulations of quasilinear theory for magnetized plasmas}}.
   \jt{Frontiers in Astronomy and Space Sciences}  \bvol{9},  \pg{1010133},
  \arxiv{arXiv: 2208.09477}.

\bibitem[Chandrasekhar {\em et~al.\/}(1958)Chandrasekhar, Kaufman \&
  Watson]{chandrasekhar1958stability}
{\sc \au{Chandrasekhar, Subrahmanyan}, \au{Kaufman, AN} \& \au{Watson, KM}}
  \yr{1958}  \at{The stability of the pinch}.  \jt{Proceedings of the Royal
  Society of London. Series A, Mathematical and Physical Sciences}  \pg{pp.
  435--455}.

\bibitem[Dodin \& Fisch(2005)]{Dodin2005}
{\sc \au{Dodin, I.~Y.} \& \au{Fisch, N.~J.}} \yr{2005}  \at{Variational
  formulation of the {G}ardner's restacking algorithm}.  \jt{Physics Letters,
  Section A: General, Atomic and Solid State Physics}  \bvol{341}~(1-4),
  \pg{187--192}.

\bibitem[Ewart {\em et~al.\/}(2023)Ewart, Nastac \&
  Schekochihin]{ewart_nastac_schekochihin_2023}
{\sc \au{Ewart, Robert}, \au{Nastac, Michael} \& \au{Schekochihin, Alexander}}
  \yr{2023}  \at{Non-thermal particle acceleration and power-law tails via
  relaxation to universal lynden-bell equilibria}.  \jt{Journal of Plasma
  Physics}  \bvol{89}~(5).

\bibitem[{Ewart} {\em et~al.\/}(2022){Ewart}, {Brown}, {Adkins} \&
  {Schekochihin}]{2022JPlPh..88e9201E}
{\sc \au{{Ewart}, R.~J.}, \au{{Brown}, A.}, \au{{Adkins}, T.} \&
  \au{{Schekochihin}, A.~A.}} \yr{2022}  \at{{Collisionless relaxation of a
  Lynden-Bell plasma}}.  \jt{Journal of Plasma Physics}  \bvol{88}~(5),
  \pg{925880501},  \arxiv{arXiv: 2201.03376}.

\bibitem[{Ewart} {\em et~al.\/}(2025){Ewart}, {Nastac}, {Bilbao}, {Silva},
  {Silva} \& {Schekochihin}]{Ewart2025}
{\sc \au{{Ewart}, Robert~J.}, \au{{Nastac}, Michael~L.}, \au{{Bilbao},
  Pablo~J.}, \au{{Silva}, Thales}, \au{{Silva}, Lu{\'\i}s~O.} \&
  \au{{Schekochihin}, Alexander~A.}} \yr{2025}  \at{{Relaxation to universal
  non-Maxwellian equilibria in a collisionless plasma}}.  \jt{Proceedings of
  the National Academy of Science}  \bvol{122}~(17),  \pg{e2417813122}.

\bibitem[{Feldman} {\em et~al.\/}(1975){Feldman}, {Asbridge}, {Bame},
  {Montgomery} \& {Gary}]{1975JGR....80.4181F}
{\sc \au{{Feldman}, W.~C.}, \au{{Asbridge}, J.~R.}, \au{{Bame}, S.~J.},
  \au{{Montgomery}, M.~D.} \& \au{{Gary}, S.~P.}} \yr{1975}  \at{{Solar wind
  electrons}}.  \jt{\jgr}  \bvol{80}~(31),  \pg{4181}.

\bibitem[Fisk \& Gloeckler(2006)]{fisk2006common}
{\sc \au{Fisk, LA} \& \au{Gloeckler, G}} \yr{2006}  \at{The common spectrum for
  accelerated ions in the quiet-time solar wind}.  \jt{The Astrophysical
  Journal Letters}  \bvol{640}~(1),  \pg{L79--L82}.

\bibitem[{Gardner}(1963)]{1963PhFl....6..839G}
{\sc \au{{Gardner}, Clifford~S.}} \yr{1963}  \at{{Bound on the Energy Available
  from a Plasma}}.  \jt{Physics of Fluids}  \bvol{6}~(6),  \pg{839--840}.

\bibitem[{Ghavamian} {\em et~al.\/}(2007){Ghavamian}, {Laming} \&
  {Rakowski}]{2007ApJ...654L..69G}
{\sc \au{{Ghavamian}, Parviz}, \au{{Laming}, J.~Martin} \& \au{{Rakowski},
  Cara~E.}} \yr{2007}  \at{{A Physical Relationship between Electron-Proton
  Temperature Equilibration and Mach Number in Fast Collisionless Shocks}}.
  \jt{\apjl}  \bvol{654}~(1),  \pg{L69--L72},  \arxiv{arXiv: astro-ph/0611306}.

\bibitem[{Hahn} \& {Savin}(2015)]{2015ApJ...809..178H}
{\sc \au{{Hahn}, M.} \& \au{{Savin}, D.~W.}} \yr{2015}  \at{{A Simple Method
  for Modeling Collision Processes in Plasmas with a Kappa Energy
  Distribution}}.  \jt{\apj}  \bvol{809}~(2),  \pg{178},  \arxiv{arXiv:
  1506.07127}.

\bibitem[{Helander}(2017)]{2017JPlPh..83d7101H}
{\sc \au{{Helander}, Per}} \yr{2017}  \at{{Available energy and ground states
  of collisionless plasmas}}.  \jt{Journal of Plasma Physics}  \bvol{83}~(4),
  \pg{715830401},  \arxiv{arXiv: 1706.05219}.

\bibitem[Kasper {\em et~al.\/}(2013)Kasper, Maruca, Stevens \&
  Zaslavsky]{kasper2013sensitive}
{\sc \au{Kasper, Justin~C}, \au{Maruca, Bennett~A}, \au{Stevens, Michael~L} \&
  \au{Zaslavsky, Arnaud}} \yr{2013}  \at{Sensitive test for ion-cyclotron
  resonant heating in the solar wind}.  \jt{Physical review letters}
  \bvol{110}~(9),  \pg{091102}.

\bibitem[{Kennel} \& {Engelmann}(1966)]{1966PhFl....9.2377K}
{\sc \au{{Kennel}, C.~F.} \& \au{{Engelmann}, F.}} \yr{1966}  \at{{Velocity
  Space Diffusion from Weak Plasma Turbulence in a Magnetic Field}}.
  \jt{Physics of Fluids}  \bvol{9}~(12),  \pg{2377--2388}.

\bibitem[{Klein} {\em et~al.\/}(2018){Klein}, {Alterman}, {Stevens}, {Vech} \&
  {Kasper}]{klein2018majority}
{\sc \au{{Klein}, K.~G.}, \au{{Alterman}, B.~L.}, \au{{Stevens}, M.~L.},
  \au{{Vech}, D.} \& \au{{Kasper}, J.~C.}} \yr{2018}  \at{{Majority of Solar
  Wind Intervals Support Ion-Driven Instabilities}}.  \jt{\prl}
  \bvol{120}~(20),  \pg{205102},  \arxiv{arXiv: 1804.06330}.

\bibitem[Klein {\em et~al.\/}(2017)Klein, Kasper, Korreck \&
  Stevens]{klein2017applying}
{\sc \au{Klein, Kristopher~G}, \au{Kasper, Justin~C}, \au{Korreck, KE} \&
  \au{Stevens, Michael~L}} \yr{2017}  \at{Applying nyquist's method for
  stability determination to solar wind observations}.  \jt{Journal of
  Geophysical Research: Space Physics}  \bvol{122}~(10),  \pg{9815--9823}.

\bibitem[Klein {\em et~al.\/}(2019)Klein, Martinovi{\'c}, Stansby \&
  Horbury]{klein2019linear}
{\sc \au{Klein, Kristopher~G}, \au{Martinovi{\'c}, Mihailo}, \au{Stansby,
  David} \& \au{Horbury, Timothy~S}} \yr{2019}  \at{Linear stability in the
  inner heliosphere: Helios re-evaluated}.  \jt{The Astrophysical Journal}
  \bvol{887}~(2),  \pg{234}.

\bibitem[Krasnoselskikh {\em et~al.\/}(2013)Krasnoselskikh, Balikhin, Walker,
  Schwartz, Sundkvist, Lobzin, Gedalin, Bale, Mozer, Soucek {\em
  et~al.\/}]{krasnoselskikh2013dynamic}
{\sc \au{Krasnoselskikh, Vladimir}, \au{Balikhin, M}, \au{Walker, SN},
  \au{Schwartz, S}, \au{Sundkvist, David}, \au{Lobzin, V}, \au{Gedalin, M},
  \au{Bale, SD}, \au{Mozer, F}, \au{Soucek, J} \& \au{others}} \yr{2013}
  \at{The dynamic quasiperpendicular shock: Cluster discoveries}.  \jt{Space
  Science Reviews}  \bvol{178}~(2),  \pg{535--598}.

\bibitem[{Lemmerz} {\em et~al.\/}(2025){Lemmerz}, {Shalaby}, {Pfrommer} \&
  {Thomas}]{2025ApJ...979...34L}
{\sc \au{{Lemmerz}, Rouven}, \au{{Shalaby}, Mohamad}, \au{{Pfrommer},
  Christoph} \& \au{{Thomas}, Timon}} \yr{2025}  \at{{The Theory of Resonant
  Cosmic Ray{\textendash}driven Instabilities{\textemdash}Growth and Saturation
  of Single Modes}}.  \jt{\apj}  \bvol{979}~(1),  \pg{34},  \arxiv{arXiv:
  2406.04400}.

\bibitem[{Livadiotis}(2018)]{2018EL....12250001L}
{\sc \au{{Livadiotis}, George}} \yr{2018}  \at{{Thermodynamic origin of kappa
  distributions}}.  \jt{EPL (Europhysics Letters)}  \bvol{122}~(5),
  \pg{50001}.

\bibitem[{Livadiotis} \& {McComas}(2025)]{2025ApJ...982..169L}
{\sc \au{{Livadiotis}, G.} \& \au{{McComas}, D.~J.}} \yr{2025}  \at{{What
  Defines Stationarity in Space Plasmas}}.  \jt{\apj}  \bvol{982}~(2),
  \pg{169},  \arxiv{arXiv: 2502.13168}.

\bibitem[Lynden-Bell(1967)]{10.1093/mnras/136.1.101}
{\sc \au{Lynden-Bell, D.}} \yr{1967}  \at{Statistical mechanics of violent
  relaxation in stellar systems}.  \jt{Monthly Notices of the Royal
  Astronomical Society}  \bvol{136}~(1),  \pg{101--121},  \arxiv{arXiv:
  https://academic.oup.com/mnras/article-pdf/136/1/101/8075239/mnras136-0101.pdf}.

\bibitem[{Lyons}(1974)]{1974JPlPh..12...45L}
{\sc \au{{Lyons}, Lawrence~R.}} \yr{1974}  \at{{General relations for resonant
  particle diffusion in pitch angle and energy}}.  \jt{Journal of Plasma
  Physics}  \bvol{12}~(1),  \pg{45--49}.

\bibitem[{Malkov} \& {Jebaraj}(2026)]{2026arXiv260109807M}
{\sc \au{{Malkov}, Mikhail} \& \au{{Jebaraj}, Immanuel}} \yr{2026}
  \at{{Magnetic Pumping: Plasma Heating to Particle Acceleration}}.  \jt{arXiv
  e-prints}  \pg{p. arXiv:2601.09807},  \arxiv{arXiv: 2601.09807}.

\bibitem[Marsch(2012)]{marsch2012helios}
{\sc \au{Marsch, Eckart}} \yr{2012}  \at{Helios: evolution of distribution
  functions 0.3--1 au}.  \jt{Space science reviews}  \bvol{172}~(1),
  \pg{23--39}.

\bibitem[Maruca {\em et~al.\/}(2012)Maruca, Kasper \&
  Gary]{maruca2012instability}
{\sc \au{Maruca, Bennett~A}, \au{Kasper, Justin~C} \& \au{Gary, S~Peter}}
  \yr{2012}  \at{Instability-driven limits on helium temperature anisotropy in
  the solar wind: observations and linear vlasov analysis}.  \jt{The
  Astrophysical Journal}  \bvol{748}~(2),  \pg{137}.

\bibitem[{Matyka} {\em et~al.\/}(2016){Matyka}, {Go{\l}embiewski} \&
  {Koza}]{2016PhRvE..93a3110M}
{\sc \au{{Matyka}, Maciej}, \au{{Go{\l}embiewski}, Jaros{\l}aw} \& \au{{Koza},
  Zbigniew}} \yr{2016}  \at{{Power-exponential velocity distributions in
  disordered porous media}}.  \jt{\pre}  \bvol{93}~(1),  \pg{013110},
  \arxiv{arXiv: 1509.04873}.

\bibitem[Nastac {\em et~al.\/}(2024)Nastac, Ewart, Sengupta, Schekochihin,
  Barnes \& Dorland]{nastac2024phase}
{\sc \au{Nastac, Michael~L}, \au{Ewart, Robert~J}, \au{Sengupta, Wrick},
  \au{Schekochihin, Alexander~A}, \au{Barnes, Michael} \& \au{Dorland,
  William~D}} \yr{2024}  \at{Phase-space entropy cascade and irreversibility of
  stochastic heating in nearly collisionless plasma turbulence}.  \jt{Physical
  Review E}  \bvol{109}~(6),  \pg{065210}.

\bibitem[{Parker}(1958)]{1958Parker}
{\sc \au{{Parker}, E.~N.}} \yr{1958}  \at{{Dynamics of the Interplanetary Gas
  and Magnetic Fields.}}  \jt{\apj}  \bvol{128},  \pg{664}.

\bibitem[Paschmann \& Daly(1998)]{Paschmann1998AnalysisMF}
{\sc \au{Paschmann, G.} \& \au{Daly, Patrick~W.}} \yr{1998} Analysis methods
  for multi-spacecraft data.  \bt{In {\em Analysis methods for multi-spacecraft
  data\/}}.

\bibitem[{Penrose}(1960)]{1960PhFl....3..258P}
{\sc \au{{Penrose}, Oliver}} \yr{1960}  \at{{Electrostatic Instabilities of a
  Uniform Non-Maxwellian Plasma}}.  \jt{Physics of Fluids}  \bvol{3}~(2),
  \pg{258--265}.

\bibitem[{Perrone} {\em et~al.\/}(2026){Perrone}, {Berlok}, {Puchwein} \&
  {Pfrommer}]{2026arXiv260106250P}
{\sc \au{{Perrone}, Lorenzo~Maria}, \au{{Berlok}, Thomas}, \au{{Puchwein},
  Ewald} \& \au{{Pfrommer}, Christoph}} \yr{2026}  \at{{Characterizing
  turbulence in galaxy clusters: defining turbulent energies and assessing
  multi-scale versus fixed-scale filters}}.  \jt{arXiv e-prints}  \pg{p.
  arXiv:2601.06250},  \arxiv{arXiv: 2601.06250}.

\bibitem[{Raymond} {\em et~al.\/}(2023){Raymond}, {Ghavamian}, {Bohdan}, {Ryu},
  {Niemiec}, {Sironi}, {Tran}, {Amato}, {Hoshino}, {Pohl}, {Amano} \&
  {Fiuza}]{2023ApJ...949...50R}
{\sc \au{{Raymond}, John~C.}, \au{{Ghavamian}, Parviz}, \au{{Bohdan}, Artem},
  \au{{Ryu}, Dongsu}, \au{{Niemiec}, Jacek}, \au{{Sironi}, Lorenzo},
  \au{{Tran}, Aaron}, \au{{Amato}, Elena}, \au{{Hoshino}, Masahiro},
  \au{{Pohl}, Martin}, \au{{Amano}, Takanobu} \& \au{{Fiuza}, Frederico}}
  \yr{2023}  \at{{Electron-Ion Temperature Ratio in Astrophysical Shocks}}.
  \jt{\apj}  \bvol{949}~(2),  \pg{50},  \arxiv{arXiv: 2303.08849}.

\bibitem[{Schekochihin} \& {Cowley}(2006)]{2006AN....327..599S}
{\sc \au{{Schekochihin}, A.~A.} \& \au{{Cowley}, S.~C.}} \yr{2006}  \at{{Fast
  growth of magnetic fields in galaxy clusters: a self-accelerating dynamo}}.
  \jt{Astronomische Nachrichten}  \bvol{327},  \pg{599},  \arxiv{arXiv:
  astro-ph/0508535}.

\bibitem[{Schekochihin} {\em et~al.\/}(2008){Schekochihin}, {Cowley},
  {Kulsrud}, {Rosin} \& {Heinemann}]{2008PhRvL.100h1301S}
{\sc \au{{Schekochihin}, A.~A.}, \au{{Cowley}, S.~C.}, \au{{Kulsrud}, R.~M.},
  \au{{Rosin}, M.~S.} \& \au{{Heinemann}, T.}} \yr{2008}  \at{{Nonlinear Growth
  of Firehose and Mirror Fluctuations in Astrophysical Plasmas}}.  \jt{\prl}
  \bvol{100}~(8),  \pg{081301},  \arxiv{arXiv: 0709.3828}.

\bibitem[{Schekochihin} {\em et~al.\/}(2010){Schekochihin}, {Cowley}, {Rincon}
  \& {Rosin}]{2010MNRAS.405..291S}
{\sc \au{{Schekochihin}, A.~A.}, \au{{Cowley}, S.~C.}, \au{{Rincon}, F.} \&
  \au{{Rosin}, M.~S.}} \yr{2010}  \at{{Magnetofluid dynamics of magnetized
  cosmic plasma: firehose and gyrothermal instabilities}}.  \jt{\mnras}
  \bvol{405}~(1),  \pg{291--300},  \arxiv{arXiv: 0912.1359}.

\bibitem[{Schlickeiser}(2002)]{2002cra..book.....S}
{\sc \au{{Schlickeiser}, Reinhard}} \yr{2002} {\em {Cosmic Ray
  Astrophysics}\/}.

\bibitem[{Shalaby} {\em et~al.\/}(2022){Shalaby}, {Lemmerz}, {Thomas} \&
  {Pfrommer}]{2022ApJ...932...86S}
{\sc \au{{Shalaby}, Mohamad}, \au{{Lemmerz}, Rouven}, \au{{Thomas}, Timon} \&
  \au{{Pfrommer}, Christoph}} \yr{2022}  \at{{The Mechanism of Efficient
  Electron Acceleration at Parallel Nonrelativistic Shocks}}.  \jt{\apj}
  \bvol{932}~(2),  \pg{86},  \arxiv{arXiv: 2202.05288}.

\bibitem[{Shalaby} {\em et~al.\/}(2021){Shalaby}, {Thomas} \&
  {Pfrommer}]{2021ApJ...908..206S}
{\sc \au{{Shalaby}, Mohamad}, \au{{Thomas}, Timon} \& \au{{Pfrommer},
  Christoph}} \yr{2021}  \at{{A New Cosmic-Ray-driven Instability}}.  \jt{\apj}
   \bvol{908}~(2),  \pg{206},  \arxiv{arXiv: 2010.11197}.

\bibitem[{Sharma} {\em et~al.\/}(2006){Sharma}, {Hammett}, {Quataert} \&
  {Stone}]{2006ApJ...637..952S}
{\sc \au{{Sharma}, Prateek}, \au{{Hammett}, Gregory~W.}, \au{{Quataert}, Eliot}
  \& \au{{Stone}, James~M.}} \yr{2006}  \at{{Shearing Box Simulations of the
  MRI in a Collisionless Plasma}}.  \jt{\apj}  \bvol{637}~(2),  \pg{952--967},
  \arxiv{arXiv: astro-ph/0508502}.

\bibitem[Southwood \& Kivelson(1993)]{southwood1993mirror}
{\sc \au{Southwood, David~J} \& \au{Kivelson, Margaret~G}} \yr{1993}
  \at{Mirror instability: 1. physical mechanism of linear instability}.
  \jt{Journal of Geophysical Research: Space Physics}  \bvol{98}~(A6),
  \pg{9181--9187}.

\bibitem[Tong {\em et~al.\/}(2019)Tong, Vasko, Artemyev, Bale \&
  Mozer]{tong2019statistical}
{\sc \au{Tong, Yuguang}, \au{Vasko, Ivan~Y}, \au{Artemyev, Anton~V}, \au{Bale,
  Stuart~D} \& \au{Mozer, Forrest~S}} \yr{2019}  \at{Statistical study of
  whistler waves in the solar wind at 1 au}.  \jt{The Astrophysical Journal}
  \bvol{878}~(1),  \pg{41}.

\bibitem[{Tran} {\em et~al.\/}(2023){Tran}, {Sironi}, {Ley}, {Zweibel} \&
  {Riquelme}]{2023ApJ...948..130T}
{\sc \au{{Tran}, Aaron}, \au{{Sironi}, Lorenzo}, \au{{Ley}, Francisco},
  \au{{Zweibel}, Ellen~G.} \& \au{{Riquelme}, Mario~A.}} \yr{2023}
  \at{{Electron Reacceleration via Ion Cyclotron Waves in the Intracluster
  Medium}}.  \jt{\apj}  \bvol{948}~(2),  \pg{130},  \arxiv{arXiv: 2209.12902}.

\bibitem[Treumann(2009)]{treumann2009fundamentals}
{\sc \au{Treumann, RA}} \yr{2009}  \at{Fundamentals of collisionless shocks for
  astrophysical application, 1. non-relativistic shocks}.  \jt{The Astronomy
  and Astrophysics Review}  \bvol{17}~(4),  \pg{409--535}.

\bibitem[Verscharen {\em et~al.\/}(2022)Verscharen, Chandran, Boella, Halekas,
  Innocenti, Jagarlamudi, Micera, Pierrard, {\v{S}}tver{\'a}k, Vasko {\em
  et~al.\/}]{verscharen2022electron}
{\sc \au{Verscharen, Daniel}, \au{Chandran, Benjamin~DG}, \au{Boella,
  Elisabetta}, \au{Halekas, Jasper}, \au{Innocenti, Maria~Elena},
  \au{Jagarlamudi, Vamsee~K}, \au{Micera, Alfredo}, \au{Pierrard, Viviane},
  \au{{\v{S}}tver{\'a}k, {\v{S}}}, \au{Vasko, Ivan~Y} \& \au{others}} \yr{2022}
   \at{Electron-driven instabilities in the solar wind}.  \jt{Frontiers in
  Astronomy and Space Sciences}  \bvol{9},  \pg{951628}.

\bibitem[{Verscharen} {\em et~al.\/}(2019){Verscharen}, {Klein} \&
  {Maruca}]{2019LRSP...16....5V}
{\sc \au{{Verscharen}, Daniel}, \au{{Klein}, Kristopher~G.} \& \au{{Maruca},
  Bennett~A.}} \yr{2019}  \at{{The multi-scale nature of the solar wind}}.
  \jt{Living Reviews in Solar Physics}  \bvol{16}~(1),  \pg{5},  \arxiv{arXiv:
  1902.03448}.

\bibitem[Villani(2006{\natexlab{{\em a\/}}})]{doi:10.1142/9789812704016-0011}
{\sc \au{Villani, C{\'e}dric}} \yr{2006{\natexlab{{\em a\/}}}} {\em Entropy
  production and convergence to equilibrium for the Boltzmann equation\/},
  \pg{pp. 130--144},  \arxiv{arXiv:
  https://www.worldscientific.com/doi/pdf/10.1142/9789812704016\_0011}.

\bibitem[Villani(2006{\natexlab{{\em b\/}}})]{villani2006mathematics}
{\sc \au{Villani, C{\'e}dric}} \yr{2006{\natexlab{{\em b\/}}}}  \at{Mathematics
  of granular materials}.  \jt{Journal of statistical physics}  \bvol{124}~(2),
   \pg{781--822}.

\bibitem[{Xrism Collaboration} {\em et~al.\/}(2025){Xrism Collaboration},
  {Audard}, {Awaki}, {Ballhausen}, {Bamba}, {Behar}, {Boissay-Malaquin},
  {Brenneman}, {Brown}, {Corrales}, {Costantini}, {Cumbee}, {Diaz Trigo},
  {Done}, {Dotani}, {Ebisawa}, {Eckart}, {Eckert}, {Eguchi}, {Enoto}, {Ezoe},
  {Foster}, {Fujimoto}, {Fujita}, {Fukazawa}, {Fukushima}, {Furuzawa}, {Gallo},
  {Garc{\'\i}a}, {Gu}, {Guainazzi}, {Hagino}, {Hamaguchi}, {Hatsukade},
  {Hayashi}, {Hayashi}, {Hell}, {Hodges-Kluck}, {Hornschemeier}, {Ichinohe},
  {Ishida}, {Ishikawa}, {Ishisaki}, {Kaastra}, {Kallman}, {Kara}, {Katsuda},
  {Kanemaru}, {Kelley}, {Kilbourne}, {Kitamoto}, {Kobayashi}, {Kohmura},
  {Kubota}, {Leutenegger}, {Loewenstein}, {Maeda}, {Markevitch}, {Matsumoto},
  {Matsushita}, {McCammon}, {McNamara}, {Mernier}, {Miller}, {Miller},
  {Mitsuishi}, {Mizumoto}, {Mizuno}, {Mori}, {Mukai}, {Murakami}, {Mushotzky},
  {Nakajima}, {Nakazawa}, {Ness}, {Nobukawa}, {Nobukawa}, {Noda}, {Odaka},
  {Ogawa}, {Ogorzalek}, {Okajima}, {Ota}, {Paltani}, {Petre}, {Plucinsky},
  {Porter}, {Pottschmidt}, {Sato}, {Sato}, {Sawada}, {Seta}, {Shidatsu},
  {Simionescu}, {Smith}, {Suzuki}, {Szymkowiak}, {Takahashi}, {Takeo},
  {Tamagawa}, {Tamura}, {Tanaka}, {Tanimoto}, {Tashiro}, {Terada}, {Terashima},
  {Tsuboi}, {Tsujimoto}, {Tsunemi}, {Tsuru}, {Uchida}, {Uchida}, {Uchida},
  {Uchiyama}, {Ueda}, {Uno}, {Vink}, {Watanabe}, {Williams}, {Yamada},
  {Yamada}, {Yamaguchi}, {Yamaoka}, {Yamasaki}, {Yamauchi}, {Yamauchi},
  {Yaqoob}, {Yoneyama}, {Yoshida}, {Yukita}, {Zhuravleva}, {Bartalesi},
  {Ettori}, {Kosarzycki}, {Lovisari}, {Rose}, {Sarkar}, {Sun} \&
  {Tamhane}]{2025ApJ...982L...5X}
{\sc \au{{Xrism Collaboration}}, \au{{Audard}, Marc}, \au{{Awaki}, Hisamitsu},
  \au{{Ballhausen}, Ralf}, \au{{Bamba}, Aya}, \au{{Behar}, Ehud},
  \au{{Boissay-Malaquin}, Rozenn}, \au{{Brenneman}, Laura}, \au{{Brown},
  Gregory~V.}, \au{{Corrales}, Lia}, \au{{Costantini}, Elisa}, \au{{Cumbee},
  Renata}, \au{{Diaz Trigo}, Maria}, \au{{Done}, Chris}, \au{{Dotani},
  Tadayasu}, \au{{Ebisawa}, Ken}, \au{{Eckart}, Megan~E.}, \au{{Eckert},
  Dominique}, \au{{Eguchi}, Satoshi}, \au{{Enoto}, Teruaki}, \au{{Ezoe},
  Yuichiro}, \au{{Foster}, Adam}, \au{{Fujimoto}, Ryuichi}, \au{{Fujita},
  Yutaka}, \au{{Fukazawa}, Yasushi}, \au{{Fukushima}, Kotaro}, \au{{Furuzawa},
  Akihiro}, \au{{Gallo}, Luigi}, \au{{Garc{\'\i}a}, Javier~A.}, \au{{Gu},
  Liyi}, \au{{Guainazzi}, Matteo}, \au{{Hagino}, Kouichi}, \au{{Hamaguchi},
  Kenji}, \au{{Hatsukade}, Isamu}, \au{{Hayashi}, Katsuhiro}, \au{{Hayashi},
  Takayuki}, \au{{Hell}, Natalie}, \au{{Hodges-Kluck}, Edmund},
  \au{{Hornschemeier}, Ann}, \au{{Ichinohe}, Yuto}, \au{{Ishida}, Manabu},
  \au{{Ishikawa}, Kumi}, \au{{Ishisaki}, Yoshitaka}, \au{{Kaastra}, Jelle},
  \au{{Kallman}, Timothy}, \au{{Kara}, Erin}, \au{{Katsuda}, Satoru},
  \au{{Kanemaru}, Yoshiaki}, \au{{Kelley}, Richard}, \au{{Kilbourne},
  Caroline}, \au{{Kitamoto}, Shunji}, \au{{Kobayashi}, Shogo}, \au{{Kohmura},
  Takayoshi}, \au{{Kubota}, Aya}, \au{{Leutenegger}, Maurice},
  \au{{Loewenstein}, Michael}, \au{{Maeda}, Yoshitomo}, \au{{Markevitch},
  Maxim}, \au{{Matsumoto}, Hironori}, \au{{Matsushita}, Kyoko}, \au{{McCammon},
  Dan}, \au{{McNamara}, Brian}, \au{{Mernier}, Fran{\c{c}}ois}, \au{{Miller},
  Eric~D.}, \au{{Miller}, Jon~M.}, \au{{Mitsuishi}, Ikuyuki}, \au{{Mizumoto},
  Misaki}, \au{{Mizuno}, Tsunefumi}, \au{{Mori}, Koji}, \au{{Mukai}, Koji},
  \au{{Murakami}, Hiroshi}, \au{{Mushotzky}, Richard}, \au{{Nakajima},
  Hiroshi}, \au{{Nakazawa}, Kazuhiro}, \au{{Ness}, Jan-Uwe}, \au{{Nobukawa},
  Kumiko}, \au{{Nobukawa}, Masayoshi}, \au{{Noda}, Hirofumi}, \au{{Odaka},
  Hirokazu}, \au{{Ogawa}, Shoji}, \au{{Ogorzalek}, Anna}, \au{{Okajima},
  Takashi}, \au{{Ota}, Naomi}, \au{{Paltani}, Stephane}, \au{{Petre}, Robert},
  \au{{Plucinsky}, Paul}, \au{{Porter}, Frederick~S.}, \au{{Pottschmidt},
  Katja}, \au{{Sato}, Kosuke}, \au{{Sato}, Toshiki}, \au{{Sawada}, Makoto},
  \au{{Seta}, Hiromi}, \au{{Shidatsu}, Megumi}, \au{{Simionescu}, Aurora},
  \au{{Smith}, Randall}, \au{{Suzuki}, Hiromasa}, \au{{Szymkowiak}, Andrew},
  \au{{Takahashi}, Hiromitsu}, \au{{Takeo}, Mai}, \au{{Tamagawa}, Toru},
  \au{{Tamura}, Keisuke}, \au{{Tanaka}, Takaaki}, \au{{Tanimoto}, Atsushi},
  \au{{Tashiro}, Makoto}, \au{{Terada}, Yukikatsu}, \au{{Terashima}, Yuichi},
  \au{{Tsuboi}, Yohko}, \au{{Tsujimoto}, Masahiro}, \au{{Tsunemi}, Hiroshi},
  \au{{Tsuru}, Takeshi}, \au{{Uchida}, Hiroyuki}, \au{{Uchida}, Nagomi},
  \au{{Uchida}, Yuusuke}, \au{{Uchiyama}, Hideki}, \au{{Ueda}, Yoshihiro},
  \au{{Uno}, Shinichiro}, \au{{Vink}, Jacco}, \au{{Watanabe}, Shin},
  \au{{Williams}, Brian~J.}, \au{{Yamada}, Satoshi}, \au{{Yamada}, Shinya},
  \au{{Yamaguchi}, Hiroya}, \au{{Yamaoka}, Kazutaka}, \au{{Yamasaki}, Noriko},
  \au{{Yamauchi}, Makoto}, \au{{Yamauchi}, Shigeo}, \au{{Yaqoob}, Tahir},
  \au{{Yoneyama}, Tomokage}, \au{{Yoshida}, Tessei}, \au{{Yukita}, Mihoko},
  \au{{Zhuravleva}, Irina}, \au{{Bartalesi}, Tommaso}, \au{{Ettori}, Stefano},
  \au{{Kosarzycki}, Roman}, \au{{Lovisari}, Lorenzo}, \au{{Rose}, Tom},
  \au{{Sarkar}, Arnab}, \au{{Sun}, Ming} \& \au{{Tamhane}, Prathamesh}}
  \yr{2025}  \at{{XRISM Reveals Low Nonthermal Pressure in the Core of the Hot,
  Relaxed Galaxy Cluster A2029}}.  \jt{\apjl}  \bvol{982}~(1),  \pg{L5},
  \arxiv{arXiv: 2501.05514}.

\bibitem[Yoon(2017)]{yoon_kinetic_2017}
{\sc \au{Yoon, Peter~H.}} \yr{2017}  \at{Kinetic instabilities in the solar
  wind driven by temperature anisotropies}.  \jt{Reviews of Modern Plasma
  Physics}  \bvol{1}~(1),  \pg{4}.

\end{thebibliography}

\end{document}